\documentclass[prd,notitlepage,nofootinbib,floats,superscriptaddress,eqsecnum,tightenlines,11pt]{revtex4-1}

\usepackage{graphicx}
\usepackage{amsmath,amssymb,amsfonts,amsthm,latexsym,stmaryrd}
\usepackage{marginnote}
\usepackage{color}

\definecolor{ultramarine}{rgb}{0.07, 0.04, 0.56}
\definecolor{cadmiumgreen}{rgb}{0.0, 0.42, 0.24}
\definecolor{indigo(dye)}{rgb}{0.0, 0.25, 0.42}
\usepackage[linktocpage=true]{hyperref}
\hypersetup{
colorlinks=true,
citecolor=ultramarine,
linkcolor=cadmiumgreen,
urlcolor=indigo(dye),
}
\def\beq{\begin{equation}}
\def\eeq{\end{equation}}
\newcommand{\bea}{\begin{eqnarray}}
\newcommand{\eea}{\end{eqnarray}}
\def\be{\begin{equation}}
\def\ee{\end{equation}}
\def\bi{\begin{itemize}}
\def\ei{\end{itemize}}
\def\ba{\begin{array}}
\def\ea{\end{array}}
\def\bfig{\begin{figure}}
\def\efig{\end{figure}}

\def\d{\delta}
\def\f{\frac}
\def\pd{\partial}
\def\pa{\partial}

\def\dq{{\dot q}}

\def\dQ{{\dot Q}}
\newcommand{\mk}[1]{\left( #1 \right)}
\newcommand{\kk}[1]{\left[ #1 \right]}
\newcommand{\ck}[1]{\left\{ #1 \right\}}
\newcommand{\emp}[1]{\par\vspace{2mm}\noindent{\bf{#1}}}

\begin{document}%%%%%%%%%%%%%%%%%%%%%%%%%%%%%%%%%%%%%%%%%

\title{Healthy degenerate theories with higher derivatives}

\begin{abstract}%%%%%%%%%%%%%%%%%%%%%%%%%%%%%%%%%%%%%%%%%
In the context of classical mechanics, we study the conditions 
under which higher-order derivative theories can evade the so-called Ostrogradsky instability.
More precisely, we consider general Lagrangians with second order time derivatives, of the form 
$L(\ddot\phi^a,\dot\phi^a,\phi^a;\dot q^i,q^i)$ with $a = 1,\cdots, n$ and $i = 1,\cdots, m$. 
For $n=1$, assuming that the $q^i$'s form a nondegenerate subsystem,
we confirm that the degeneracy of the kinetic matrix eliminates the Ostrogradsky instability.
The degeneracy implies,  in the Hamiltonian formulation of the theory,  the existence of a primary constraint, which generates a secondary constraint, thus eliminating the Ostrogradsky ghost.
For $n>1$, we show that, in addition to the degeneracy of the kinetic matrix, one needs to impose extra conditions to ensure the presence of  a sufficient number of secondary constraints that can eliminate all the Ostrogradsky ghosts.  When these conditions that ensure the disappearance of the Ostrogradsky instability are satisfied, we show that the  Euler-Lagrange equations, which involve a priori higher order derivatives, can be reduced to a second order system. 
\end{abstract}%%%%%%%%%%%%%%%%%%%%%%%%%%%%%%%%%%%%%%%%%

\author{Hayato Motohashi}
\email{motohashi@kicp.uchicago.edu}
\affiliation{Kavli Institute for Cosmological Physics, The University of Chicago, 
Chicago, Illinois 60637, U.S.A.}
\author{Karim Noui}
\email{karim.noui@lmpt.univ-tours.fr}
\affiliation{Laboratoire de Math\'ematiques et Physique Th\'eorique, Universit\'e Fran\c cois Rabelais, Parc de Grandmont, 37200 Tours, France}
\affiliation{Laboratoire APC -- Astroparticule et Cosmologie, Universit\'e Paris Diderot Paris 7, 75013 Paris, France}
\author{Teruaki Suyama}
\email{suyama@resceu.s.u-tokyo.ac.jp}
\affiliation{Research Center for the Early Universe (RESCEU),
Graduate School of Science,
The University of Tokyo, Tokyo 113-0033, Japan}
\author{Masahide Yamaguchi}
\email{gucci@phys.titech.ac.jp}
\affiliation{Department of Physics, Tokyo Institute of Technology,
2-12-1 Ookayama, Meguro-ku, Tokyo 152-8551, Japan}
\author{David  Langlois}
\email{langlois@apc.univ-paris7.fr}
\affiliation{Laboratoire APC -- Astroparticule et Cosmologie, Universit\'e Paris Diderot Paris 7, 75013 Paris, France}

\maketitle

\section{Introduction}%%%%%%%%%%%%%%%%%%%%%%%%%%%%%%%%%%%%%%%%%

While present observations are compatible with general relativity with a cosmological constant, it is of interest to explore  alternative theories as they could provide a more fundamental description of present data  or better account for future observations. Among these alternative theories, special attention has been devoted to Horndeski theory~\cite{Horndeski:1974wa}, or generalized Galileon theory~\cite{Nicolis:2008in,Deffayet:2009wt,Deffayet:2009mn,Deffayet:2011gz,Kobayashi:2011nu}, defined by  the most general scalar-tensor Lagrangian that yields second-order Euler-Lagrange equations of motion.  Recently it has been pointed out that one can find healthy extensions of  Horndeski theories~\cite{Gleyzes:2014dya,Gleyzes:2014qga}, whose Euler-Lagrange equations involve higher order derivatives (see also \cite{Zumalacarregui:2013pma} for an earlier example  of theory ``beyond Horndeski'' ).

To construct a sensible theory with higher order derivatives, one needs to avoid the presence of additional degrees of freedom (DOF) causing an instability due to the linear dependence of the Hamiltonian on momenta, known as the Ostrogradsky instability~\cite{Ostrogradsky,Woodard:2006nt,Woodard:2015zca}. This instability is inevitable for nondegenerate Lagrangians, unless one introduces ``by hand'' additional constraints in order to reduce the phase space~\cite{Chen:2012au}.   
Otherwise, one must turn to degenerate Lagrangians to find viable theories. Particularly interesting are Lagrangians whose degeneracy is due to the coupling between a {\it special} variable, by which we mean a variable associated with higher order derivatives in the Lagrangian, and {\it regular} variables~\cite{Langlois:2015cwa}.  
The degeneracy implies the existence of  a primary constraint and of an associated secondary constraint,  which reduce the effective dimension of phase space and thus eliminate the extra DOF at the source of the Ostrogradsky instability.  
This can be seen explicitly with the Hamiltonian analysis which has been performed recently~\cite{Langlois:2015skt} for quadratic\footnote{i.e.\ whose action depends  quadratically on the second derivatives of the scalar field.} degenerate higher order scalar tensor  (DHOST) theories introduced in \cite{Langlois:2015cwa} 
(see also \cite{Crisostomi:2016tcp,Crisostomi:2016czh,Achour:2016rkg} for subsequent works using the same approach).

Given these results, it is worth investigating how, for more general Lagrangians, the degeneracy of the kinetic matrix is related to the  elimination of the Ostrogradsky ghosts. In contrast to naive expectations, maximal degeneracy of the kinetic matrix, i.e.\ of order $n$ when  $n>1$ special variables are present, is not sufficient to eliminate multiple Ostrogradsky ghosts.  Indeed, it was found in \cite{Motohashi:2014opa} that there exist (multi-)degenerate theories involving multiple special variables which still suffer from Ostrogradsky instability as their  Hamiltonian  depends linearly on some momenta.

One of the main goals of this paper is to present, for systems with multiple special variables, a set of extra conditions, in addition to the multi-degeneracy, leading to the elimination of all Ostrogradsky ghosts. For pedagogical reasons, we consider Lagrangians that contain one special variable, then several special variables, coupled to one or several regular variables. In each case, we 
analyse these theories both from the Lagrangian and Hamiltonian points of view. 
In particular, the Hamiltonian analysis enables us to show that, while the primary constraints are directly related to the degeneracy of the Lagrangian,   the elimination of the Ostrogradsky ghosts
requires the presence of  secondary constraints. When several primary constraints are present, the number of associated secondary constraints can be lower. To eliminate all unwanted DOF associated with the linear instabilities of the Hamiltonian, one needs as many secondary constraints as primary ones, which requires some additional conditions. 
We explicitly write down how these extra conditions are expressed in the  Lagrangian formulation.  We also demonstrate how to reduce the Euler-Lagrange equations involving higher order derivatives into a second order system.

The organization of the paper is as follows.  
In Sec.~\ref{sec:single-single}, we consider  Lagrangians $L(\ddot \phi,\dot \phi, \phi ; \dot q, q)$ that depend on two variables: $\phi(t)$ with at most second order derivative, and $q(t)$ with at most first order derivative (throughout the paper, $q$ denotes a  ``regular'' variable while $\phi$ corresponds to a ``special'' variable).  We explain how  the Ostrogradsky ghost is eliminated by the constraints in the Hamiltonian formulation, and we also show that the Euler-Lagrange equations can be rewritten as a second-order system.
In Sec.~\ref{sec:single-multi}, we consider a Lagrangian $L(\ddot \phi,\dot \phi, \phi ; \dot q^i, q^i)$ with multiple regular variables $q^i$ and generalize the analysis of the previous section.  
After discussing the  case of Lagrangians $L(\ddot \phi^a,\dot \phi^a, \phi^a)$, depending only on special variables, in Sec.~\ref{sec:special-only}, we consider in Sec.~\ref{sec:multi-multi}  the most general case of  a Lagrangian $L(\ddot \phi^a,\dot \phi^a, \phi^a ; \dot q^i, q^i)$, depending on  multiple special variables $\phi^a$ and multiple regular variables.  In addition to the  degeneracy condition,  we impose extra conditions that guarantee the absence of the Ostrogradsky instability. We present the Lagrangian form of these additional conditions and show that they allow the Euler-Lagrange equations to be rewritten as a second-order system.
Sec.~\ref{sec:conc} is devoted to our conclusions.

\section{Lagrangians with single regular and special variables}%%%%%%%%%%%%%%%%%%%%%%%%%%%%%%%%%%%%%%%%%
\label{sec:single-single}

In this section, we consider a Lagrangian of the form
\bea\label{Lagrangian}
L(\ddot \phi,\dot \phi , \phi ; \dot q,q)\,
\eea 
which depends on two time-dependent variables $\phi(t)$ and $q(t)$. The Lagrangian contains at most second order derivatives of $\phi$ whereas
$q$ appears at most with first order derivatives. 
In general, theories of this type  are known to exhibit an Ostrogradsky instability. 
More precisely, 
$\phi$ and $q$ generically obey fourth order and second order equations of motion, respectively\footnote{The equation of motion for $\phi$ could
involve also third derivative of $q$. Such a third derivative can be eliminated making use of the equation of motion for $q$. Thus the dynamics is described by a system 
which is at most second order in $q$ and can be up to fourth order in $\phi$.}, 
which require 6 initial conditions in total, corresponding to three physical DOF. 
One of these DOF is a ghost with an energy unbounded from below.

Let us now derive the conditions to escape such an instability even when the Lagrangian features nontrivial second order derivatives (i.e.\ which cannot be  eliminated by simply integrating by parts). We start with a Hamiltonian analysis and find necessary and sufficient conditions for the theory to be ghost free.
To perform the Hamiltonian analysis, we consider the following equivalent Lagrangians
\be L_{eq}^{(1)} \equiv  L(\dot Q, Q, \phi ; \dot q , q) + \lambda (\dot \phi - Q) , \label{Leq1} \ee
and 
\be L_{eq}^{(2)} \equiv  L(Q_2,Q_1,\phi;Q_3,q) + \lambda_1 (\dot \phi - Q_1) + \lambda_2 (\dot Q_1-Q_2) + \lambda_3 (\dot q -Q_3), \label{Leq2}  \ee
where the $Q$'s are auxiliary variables and the $\lambda$'s are Lagrange multipliers whose associated equations of motion impose that the  $Q$'s are fixed in terms of 
$\dot q$, $\dot \phi$ and $\ddot \phi$.  
The two forms are obviously equivalent but they possess different advantages for the Hamiltonian analysis.

First,  in Sec.~\ref{ssec:ssHam1}-\ref{ssec:ssdeg}, we investigate the form \eqref{Leq1}, in which the canonical momenta capture the structure of the highest derivatives in the Lagrangian. This form is useful  to understand the physical meaning of the degeneracy of a theory in terms of additional primary constraints between the momenta.
Next, in Sec.~\ref{ssec:ssHam2},  we  consider the form \eqref{Leq2}, which is easier to generalize to a Lagrangian with an arbitrary number of variables or of derivatives, by introducing enough auxiliary variables and Lagrange multipliers so that the final Lagrangian is  of the form \eqref{Leq2} where the velocity terms only appear linearly.
Then, in Sec.~\ref{ssec:ssELeq},  we write the Euler-Lagrange equations of motion which are a priori higher order. As expected, we show that, when the theory satisfies the degeneracy condition, the equations of motion can be reduced to a second order system. 
We provide several specific examples of healthy degenerate Lagrangians in Sec.~\ref{ssec:ssexam}.

\subsection{Hamiltonian analysis}%%%%%%%%%%%%%%%%%%%%
\label{ssec:ssHam1}

Starting from the Lagrangian \eqref{Leq1}, which  depends on four variables, 
we introduce the four  canonical momenta $P$, $p$, $\pi$ and $\rho$, associated with  $Q$, $q$, $\phi$ and $\lambda$, respectively.  The (only nonvanishing) elementary Poisson brackets are thus defined by
\bea\label{Poissonbracket}
\{ Q,P\} \; = \; \{q,p \} \; = \; \{\phi,\pi \} \; = \; \{\lambda,\rho \}  \; = \; 1 \, .
\eea
From \eqref{Leq1}, it is easy to get
\bea \label{momenta}
P=\frac{\partial L}{\partial \dot Q} \equiv L_{\dot Q} \;\;, \;\;
p=\frac{\partial L}{\partial \dot q} \equiv L_{\dot q}\;\;, \;\;
\pi = \frac{\partial L_{eq}^{(1)}}{\partial \dot \phi} = \lambda \;\;, \;\;
\rho = \frac{\partial L_{eq}^{(1)}}{\partial \dot \lambda} = 0 \,.
\eea
We thus have two primary constraints
\beq \label{primary0}
\Phi = \pi -\lambda \approx 0 \;\;, \;\; 
\Psi = \rho \approx 0 \, , \eeq
whose Poisson bracket is nonvanishing: $\{ \Phi, \Psi \} = -1$.

When one can invert (at least locally in the vicinity of any point in phase space)  the first two expressions in \eqref{momenta} to obtain $\dot Q$ and $\dot q$ in terms of the momenta, then there is no further primary constraint.  Considering infinitesimal variations of the momenta with respect to $\dot Q$ and $\dot q$, we can write
\beq \label{dpddq}
\begin{pmatrix}
\d P \\
\d p
\end{pmatrix}
= K
\begin{pmatrix}
\d \dot Q \\
\d \dot q
\end{pmatrix}  .
\eeq
where the matrix $K$, which we will call  the kinetic matrix, is given by
\bea \label{kinmat}
K \equiv \left( 
\begin{array}{cc}
L_{\dot Q \dot Q} & L_{\dot q \dot Q} \\
L_{\dot q \dot Q} & L_{\dot q \dot q}
\end{array}
\right) \,\;\; \text{with the notation} \;\;\; L_{xy} \equiv \frac{\partial^2 L}{\partial x \partial y} \,.
\eea
It is possible to invert \eqref{dpddq} if and only if $K$ is nondegenerate, i.e.\ $\det K \neq 0$.

Taking into account the primary constraints (\ref{primary0}), the Hamiltonian 
can be written as
\bea
\label{Legendre}
H = H_0 + \pi Q \;\;\; \text{with} \;\;\; H_0=P\dot Q + p \dot q - L(\dot Q, Q , \phi ; \dot q , q) \, , 
\eea
where the velocities $\dot Q$ and $\dot q$ are expressed in terms of the
momenta so that $H$ depends only on the conjugate variables
(\ref{Poissonbracket}).
For the nondegenerate case, (\ref{primary0}) are the only primary constraints and the total Hamiltonian is given by
\beq \label{totalH} H_T = H(P,p,\pi,Q,q,\phi) + \mu\Phi + \nu\Psi\,, \eeq
where $\mu$ and $\nu$ are Lagrange multipliers.

Requiring stability under time evolution (with respect to the total Hamiltonian)  
of the primary constraints leads to fixing  the Lagrange multipliers to $\mu=- \{ \Psi, H \}=0$ and $\nu=\{ \Phi, H \}$. As a consequence,  there are no secondary constraints.
Therefore, we have 2 second class primary constraints which eliminate 2 initial conditions, and we end up with $(8-2)/2=3$ DOF, one of which being ghost-like.
Indeed, one sees immediately that $H$ is linear in $\pi$ which shows that,
without extra constraints, the theory features an instability. This is an illustration of the well-known Ostrogradsky instability.

\subsection{Additional primary constraint}%%%%%%%%%%%%%%%%%%%%
\label{ssec:ssadpri}

The only hope to eliminate such an instability is the existence of an additional primary constraint. Thus, we will assume, from now on,  that the momenta $P$ and $p$ are not independent variables but, instead, satisfy a relation ${\cal  R}(P,p;Q,\phi,q)=0$. In general, such a relation defines implicitly $P$ in terms of $p$ (or the reverse) 
and it is not always possible to express uniquely and globally $P$ as a function of $p$ (or the reverse). However, locally,  it is always 
possible to write either  $P = F(p,Q,\phi,q)$ or $ p= G(P,Q,\phi,q)$.
In the first case, the theory admits the additional primary constraint
\beq \label{case1}
\Xi \equiv P - F(p,Q,\phi,q) \approx 0 \,,
\eeq
whereas in the second case, the additional primary constraint is 
\beq \label{case2} \Pi \equiv p - G(P,Q,\phi,q) \approx 0 \,.  \eeq
As long as the analysis is local  in phase space,  \eqref{case1} and \eqref{case2} are equivalent, except  if $F$ (resp.\ $G$) does not depend explicitly on $p$ (resp.\ $P$) in some open set of  the phase space. 
Thus  there exists an independent case with the two unrelated primary constraint
\beq \label{case3} \tilde \Xi \equiv P - f(Q, \phi, q) \approx 0, \qquad \tilde \Pi \equiv p - g(Q, \phi, q) \approx 0\,,  \eeq 
where $f$ and $g$ depend on the coordinates only and not on the momenta.

Let us investigate the Hamiltonian structure with,  say,  the primary constraint  \eqref{case1}, in addition to the two primary constraints \eqref{primary0}.  The total Hamiltonian is now given by 
\beq H_T = H(P,p,\pi,Q,q,\phi) + \mu\Phi + \nu\Psi + \xi\Xi\,, \label{HamT}  \eeq
where $\xi$ is a new Lagrange multiplier.
Using the  Poisson bracket $\{ \Phi,\Xi \}=F_\phi$, the invariance under time evolution of  the 3 primary constraints gives
\bea
&&\dot \Phi = \{ \Phi, H \} -\nu+\xi F_\phi \approx 0, \notag\\
&&\dot \Psi = \mu \approx 0 , \\
&&\dot \Xi = \{ \Xi,H \}-\mu F_\phi \approx 0 . \notag
\eea
The first two equations enable us to  fix the two  Lagrange multipliers $\mu$ and $\nu$. Since $\mu$ is required to vanish, the third equation implies $\Theta \equiv \{ \Xi,H \} \approx 0$.
A direct computation shows that $\Theta$ is given by
\be\label{secondTheta}
\Theta=-\pi + \{ \Xi, H_0\} -F_\phi \, Q\,,
\ee
where $H_0=H-\pi Q$ has been defined in \eqref{Legendre}. A more explicit expression of $\Theta$ is given in Appendix~\ref{sec:delta}.
The  condition ${ \Theta} \approx 0$ provides a 
new constraint, independent of the previous ones, which determines $\pi$ in terms of the other phase space variables.  An important consequence is that one can get rid of the linear dependence of  the Hamiltonian $H$ on $\pi$, which signals that the Ostrogradsky instability is not present. Note however that  the Hamiltonian could  still be  unbounded from below, but for other reasons.

Let us continue our Hamiltonian analysis by requiring the time invariance of $\Theta$:
\be
{\dot \Theta} = \{ \Theta, H \}+\xi \{ \Theta, \Xi \} \approx 0,  \label{dotTheta}
\ee
where we have set $\mu \approx 0$.
For the generic case where $\Delta \equiv \{ \Theta, \Xi \}\neq 0$, the above equation  fixes $\xi$ and we have thus determined all  Lagrange multipliers.
In this case,  the theory admits 4 constraints denoted generically $\chi_i  \in (\Phi,\Psi,\Xi,\Theta)$ for $i=1,\cdots,4$. 
The Dirac matrix
\bea
D = 
\left(
\begin{array}{cccc}
0 & -1 & F_\phi & \{\Phi,\Theta \} \\
1 & 0 & 0 & 0 \\
-F_\phi & 0 & 0 & - \Delta \\
\{\Theta,\Phi\} & 0 & \Delta & 0
\end{array}
\right)
\eea
whose entries $D_{ij}=\{ \chi_i, \chi_j \} $ are the Poisson brackets between the constraints  is invertible
as $\vert \det \{ \chi_i, \chi_j \} \vert = \Delta^2$.  Thus the constraints are all second-class constraints and we end up with $(8-4)/2=2$ DOF. In  phase space, these DOF are described by the pairs $(\phi,Q)$ and $(q,p)$, as the momenta $P$ and $\pi$ have been eliminated by solving explicitly the constraints $\Xi \approx 0$ and $\Theta \approx 0$.

In the special case $\Delta= 0$, the conclusions are different and the theory possesses fewer than 2 physical DOF. In fact, two different scenarios can be encountered, depending  on whether or not  there exists a tertiary constraint. 
A tertiary constraint arises when $\Gamma \equiv \{ \Theta, H \}$ does not vanish automatically and one needs to impose it as a new constraint $\Gamma\approx 0$. 
One then has to continue the procedure and check whether $\dot \Gamma \approx 0$ generates further constraints.  Whatever the conclusion of the constraints analysis is, we are left at the end with strictly less than 2 DOF. 
In the second type of scenario, $\Gamma$ automatically vanishes and we are thus left with the four constraints $\chi_i \approx 0$. Since the skew-symmetric Dirac matrix $D$ is degenerate and contains the nonzero entry $\{\Phi,\Psi\}=-1$, one infers that it is of rank 2, which means that one can identify two first-class constraints among the four constraints. As a consequence,
we end up with only $(8-2\times 2- 2)/2= 1$ DOF in that case.

The Hamiltonian analysis of a theory with a primary constraint of the type \eqref{case2} can be performed exactly in the same way. The conclusions are strictly similar and we end up generically with 2 DOF corresponding to the  case $\Delta \neq 0$.  
As for the case \eqref{case3}, the  analysis is a bit different, as  we start with the  two primary constraints $\tilde\Xi$ and $\tilde\Pi$. 
The total Hamiltonian is then of the form
\be
H_T=H + \mu \Phi + \nu \Psi + \xi \tilde\Xi + \zeta \tilde\Pi \,,
\ee
where we have introduced the two Lagrange multipliers $\xi$ and $\zeta$.
Time invariance of these constraints fixes $\xi$ and $\zeta$, provided $\tilde \Delta \equiv \{ \tilde \Xi, \tilde \Pi \} = L_{\dot q Q} - L_{q \dot Q} \neq 0$. 
In this generic situation, we close the canonical analysis with 4 constraints $\tilde \chi_i=(\Phi,\Psi,\tilde \Xi,\tilde \Pi)$ and a  Dirac matrix which 
is invertible, since  $\vert \det \{ \tilde \chi_i, \tilde \chi_j \} \vert = \tilde \Delta^2$. We conclude that all the constraints are second class and the system contains 2 DOF. In the special situation where $\tilde \Delta = 0$, we may have more constraints or some of the constraints may become first class. In both cases, the theory possesses 1 or zero DOF.

We conclude that an additional primary constraint leads to the elimination of the unwanted ghost-like DOF.

\subsection{Degenerate Lagrangians}%%%%%%%%%%%%%%%%%%%%
\label{ssec:ssdeg}

In the previous subsection, we have assumed the existence of  explicit relations between the momenta, of the form \eqref{case1}, \eqref{case2} or \eqref{case3}, which are valid locally. A more intrinsic characterization of the corresponding Lagrangians is that their kinetic matrix, defined in \eqref{kinmat}, is degenerate.

It is immediate to check that each of the conditions \eqref{case1}--\eqref{case3}
implies the degeneracy of the kinetic matrix.
Indeed, in the case \eqref{case1} for instance, we have the relations 
\bea
L_{\dot Q \dot Q} = L_{\dot Q \dot q} F'(L_{\dot q}) \;\;\; \text{and} \;\;\; 
L_{\dot Q \dot q} = L_{\dot q \dot q} F'(L_{\dot q}) \,,
\eea
which implies immediately that the determinant of the kinetic matrix vanishes:
\beq \label{detk0} \det K = L_{\dot Q \dot Q}L_{\dot q \dot q} - L_{\dot q \dot Q} ^2 = 0 \, . \eeq
For the case \eqref{case2}, the same result holds with the replacements $q \leftrightarrow Q$ and $F\to G$.
Finally, in the case \eqref{case3}, the full kinetic matrix vanishes $K=0$, not only its determinant.

Conversely, let us now show that $\det K=0$ implies the existence of  a primary constraint of the form \eqref{case1} or \eqref{case2},  or a set of two
primary constraints  \eqref{case3}.  
First, let us consider the Lagrangians for which
\bea
L_{\dot q \dot q} = \frac{\partial p}{\partial \dot q} \neq 0\,.
\eea 
According to the implicit functions theorem, one can find  locally (in the vicinity
of any point in phase space)  a function $\varphi$ such that 
\bea
\dot q=\varphi(p,\dot Q , Q,q,\phi).
\eea 
Consequently, the momentum $P$ which depends a priori on
the two velocities ($\dot q$,$\dot Q$) and on the coordinates $(Q,q,\phi)$ can locally be expressed as a function $P=F(p,\dot Q, Q,q,\phi)$ replacing $\dot q$ by $\varphi$. 
Furthermore, the degeneracy of $K$ implies $\partial F/\partial \dot Q|_{p} = 0$. Indeed, if   $\partial F/\partial \dot Q|_{p} \neq 0$,  one could invoke the implicit functions theorem again and deduce that $\dot Q$ can be expressed in terms of the 
momenta $(P,p)$ and the coordinates, which is in contradiction with the degeneracy of $K$.  We thus conclude
\bea
P=F(p, Q,q,\phi) \,,
\eea
which corresponds precisely to the primary constraint \eqref{case1}. To summarize, $\det K =0$ together with $L_{\dot q \dot q}\neq 0$ implies that there exists a function $F(p, Q,q,\phi) $ 
such that $P=F(p, Q,q,\phi) $.

If $L_{\dot Q \dot Q} \neq 0$, a very similar analysis enables us to conclude that there exists 
now a function $G(P, Q,q,\phi)$ such that {$p=G(P, Q,q,\phi)$} 
and we recover the primary constraint \eqref{case2}.  
Note that when $\partial F/\partial p \neq 0$, then necessarily $\partial G/\partial P \neq 0$ and the two constraints are locally equivalent.
These first two cases apply to degenerate
kinetic matrix $K$ which admits only one vanishing eigenvalue.

To complete the proof, one must finally consider the cases for which  $L_{\dot q \dot q}=L_{\dot Q \dot Q}=0$. Since  the matrix $K$ is degenerate, this implies that  $K$ in fact vanishes. It is then straightforward  to show that there exist two functions $f(Q,q,\phi)$ and $g(Q,q,\phi)$ such that the constraints \eqref{case3} hold. 
A more explicit proof of this property is given  in the Appendix~\ref{sec:explicit degeneracy} without referring to the abstract implicit functions theorem.
We can derive the relations between the functions $F$, $G$ and the initial Lagrangian $L$ via this  explicit proof.

In summary,  we conclude that the condition $\det K = 0$ is equivalent to the existence of  primary constraints restricting the momenta. 
Depending on the dimension (one or two) of  the kernel of $K$, the theory admits one or two primary constraints.  It amounts to the case \eqref{case1} for $L_{\dot q\dot q}\neq 0$, the case \eqref{case2} for $L_{\dot Q\dot Q}\neq 0$, and the case \eqref{case3} for $L_{\dot Q\dot Q}=L_{\dot q\dot q}=0$.  As we said previously, the constraints
\eqref{case1} and \eqref{case2} are equivalent when $\partial F/\partial p \neq 0$ or $\partial G/\partial P \neq 0$. 
In practice, one can check whether a given Lagrangian \eqref{Leq1} is (Ostrogradsky) ghost-free by using the degeneracy condition $\det K=0$.  Then one can see whether  it has 2 DOF or less by checking $\Delta\neq 0$ or $\tilde \Delta\neq 0$.

\subsection{Alternative Hamiltonian analysis}%%%%%%%%%%%%%%%%%%%%
\label{ssec:ssHam2}

For completeness, we now perform  the Hamiltonian analysis of the Lagrangian  \eqref{Leq2}
\beq L_{eq}^{(2)} \equiv  L(Q_2,Q_1,\phi;Q_3,q) + \lambda_1 (\dot \phi - Q_1) + \lambda_2 (\dot Q_1-Q_2) + \lambda_3 (\dot q -Q_3),   \eeq
 which is equivalent to \eqref{Leq1}.
Starting with such a formulation has disadvantages. First the canonical analysis involves more constraints and thus it could be a priori 
more complicated than the analysis of \eqref{Leq1}. 
Second this formulation is more difficult to generalize and to adapt to field theories, including scalar-tensor theories. 
However, there are important benefits by considering \eqref{Leq2} in the context of this article. 
As we will see, the total Hamiltonian is explicitly defined as a function of the phase space
variables, and thus there is no need 
to resort to a local analysis to write the Hamiltonian and the constraints.
Another benefit is that we can always reduce any Lagrangian with arbitrary higher derivatives to the form \eqref{Leq2} where the velocity terms only appear linearly.

Let us start with \eqref{Leq2}. The form of the Lagrangian implies that  there are initially 8 pairs of conjugate variables
\be
\{ Q_i,P_i\} \; = \; \{\lambda_i,\rho_i \}  \; = \; \{q,p \} \; = \; \{\phi,\pi \} \; = \; 1 \,, 
\ee
with $i \in \{1,2,3\}$. It is immediate to see that we have the following 8 primary constraints 
\bea
&&\Phi_1 = \pi -\lambda_1 \approx 0 \;\;, \;\; 
\Phi_2 = P_1 - \lambda_2 \approx 0 \; \; , \;\;
\Phi_3= p - \lambda_3  \approx 0\;\; ,\;\; \cr 
&&\rho_1 \approx 0  \;\;, \;\; \rho_2 \approx 0  \;\;, \;\; \rho_3 \approx 0  \;\;, \;\;  P_2 \approx 0 \;\; \text{and} \;\; P_3 \approx 0 \,.
 \eea
Contrary to the previous subsection, {the Hamiltonian and the total Hamiltonian are now defined globally and, after simple calculations, one obtains 
\bea
&& H = - L(Q_2,Q_1,\phi;Q_3,q) + \pi Q_1 + P_1 Q_2 + p Q_3, \\
&& H_T = H +  \sum_{i=1}^3 (\mu_i \Phi_i + \nu_i \rho_i) + \xi_2 P_2 + \xi_3 P_3, 
\eea}
where $\mu_i$, $\nu_i$, $\xi_2$ and $\xi_3$ are Lagrange multipliers enforcing the primary constraints.

To pursue the canonical analysis, we compute the time evolution of the constraints and impose their conservation. The simple property
\bea
\{ \rho_i , \Phi_j\} =  \delta_{ij}
\eea
implies immediately that time conservation of the six constraints $\Phi_i \approx 0$ and $\rho_i \approx 0$ fix the Lagrange multipliers $\mu_i$ and $\nu_i$.
Thus, these primary constraints do  not generate secondary constraints. This is not the case for $P_2 \approx 0$ and $P_3 \approx 0$. Indeed, computing their time derivatives, we
obtain two  secondary constraints:
\bea  
\chi_2 &\equiv &\dot P_2 = \{ P_2 , H_T \} = L_{Q_2} - P_1 \approx 0,\\
\chi_3 &\equiv &\dot P_3 = \{ P_3 , H_T \} = L_{Q_3} - p \approx 0 \; .
\eea
These constraints are easily interpreted. They simply mean that the momentum $P_1\approx \lambda_2$ conjugate to $Q_1=\dot \phi$ is $\partial L/\partial \dot Q_1$ and
the momentum $p \approx \lambda_3$ conjugate to $q$ is $\partial L/\partial \dot q$, as expected.

We continue the analysis by computing the time evolution of these two secondary constraints and we obtain the two conditions
\bea
\dot \chi_2 & = & \{ \chi_2 , H_0 \} + L_{Q_2 Q_2} \xi_2 + L_{Q_2 Q_3} \xi_3 \approx 0, \\
\dot \chi_3 & = & \{ \chi_3 , H_0 \} + L_{Q_3 Q_2} \xi_2 + L_{Q_3 Q_3} \xi_3 \approx 0\, .
\eea
To simplify notations we have introduced  $H_0=H_T -  (\xi_2 P_2 + \xi_3 P_3)$. We note that the kinetic matrix $K$ naturally arises here when one
identifies $Q_2$ to $\dot Q=\ddot \phi$ and $Q_3$ to $\dot q$ as it should be. As a consequence, the two previous conditions can be reformulated as follows
\bea\label{conditions for xi}
K \left( \begin{array}{c}  \xi_2 \\ \xi_3 \end{array} \right) 
= \left( \begin{array}{c}  \ck{ H, \chi_2 } \\ \ck{ H, \chi_3 } \end{array} \right)\,.
\eea
The end of the analysis depends on the rank of the matrix $K$.

If $K$ is invertible, the system of equations fixes the Lagrange multipliers $\xi_2$ and $\xi_3$ and there is no further constraint. It is easy to
 check  that all the constraints are  second class.  As a consequence,
we end up with 10 second class constraints for 8 initial pairs of conjugate variables. This leads to $(16-10)/2=3$ DOF, which  include  the Ostrogradsky ghost.

If the kernel of $K$ is one-dimensional,  in the direction  $(u^2,u^3)$, one obtains  the tertiary constraint 
\bea
\Xi \equiv u^2 \{H_0,\chi_2\} + u^3 \{H_0,\chi_3\} \,,
\eea
where $u^2$ and $u^3$ are  functions of $(Q_i,q,\phi)$. This constraint is the analog of \eqref{case1}
or \eqref{case2} in the previous analysis. Requiring  time invariance of this constraint generically gives one additional  constraint, which leads to a total of 2 DOF.

Finally, when $K$ vanishes, \eqref{conditions for xi} implies two constraints
\bea
 \{ \chi_2 , H \}  \approx 0 \;\;\; \text{and} \;\;\;  \{ \chi_3 , H \} \approx 0,
\eea 
which are the analog of \eqref{case3} in the previous analysis. The discussion of this case is similar to that of \eqref{case3} and we end up in general with 2 DOF (or less).

In conclusion, we have checked that the two analyses starting from the Lagrangians \eqref{Leq1} or \eqref{Leq2} are  equivalent.

\subsection{Euler-Lagrange equations}%%%%%%%%%%%%%%%%%%%%
\label{ssec:ssELeq}

We now proceed to study the equations of motion in presence of  either of the primary constraints \eqref{case1}--\eqref{case3}. For a general Lagrangian of the form (\ref{Lagrangian}), the Euler-Lagrange equations read
\bea
&&L_\phi - \frac{d L_{\dot \phi}}{d t} + \frac{d^2 L_{\ddot \phi}}{dt^2}  =  0, \label{eom1}\\
&&L_q - \frac{d L_{\dot q}}{d t}  =  0 \,.\label{eom2}
\eea
Due to the dependence of $L$ on $\ddot\phi$, the equation of motion for $q$ (\ref{eom2})
in general involves  the third derivative of  $\phi$. As for the equation of motion for $\phi$ (\ref{eom1}), it
involves the fourth derivative of $\phi$ and the third derivative of $q$.

When the theory is degenerate,  the  equations of motion can be reformulated as a second order system, as we now show.   In order to make the correspondence with the Hamiltonian analysis clearer, we first replace \eqref{eom1} and \eqref{eom2}  by the equivalent Euler-Lagrange equations  derived from
the alternative Lagrangian \eqref{Leq1}:
\bea \label{EOMLeq1}
L_Q - \frac{d L_{\dot Q}}{dt} = \lambda \; , \qquad 
L_q - \frac{d L_{\dot q}}{dt}= 0 \; , \qquad  L_\phi = \dot \lambda \;,\qquad \dot \phi = Q \, .
\eea
Let us concentrate on the first two equations which can easily be  rewritten in a more explicit way as
\bea \label{system1}
K \left( \begin{array}{c} \ddot Q \\ \ddot q \end{array}\right) = 
\left( \begin{array}{c} V - \lambda \\ v \end{array}\right) \qquad \text{with} \;\; 
\left( \begin{array}{c} V \\ v \end{array}\right)=
\left( \begin{array}{c}  L_Q -L_{\dot Q Q} \dot Q -L_{\dot Q q} \dot q - L_{\dot Q \phi} \dot \phi   \\  
 L_q  -L_{\dot q Q} \dot Q - L_{\dot q q} \dot q - L_{\dot q \phi} \dot \phi \end{array}\right)\,.
\eea
As the kinetic matrix $K$ is degenerate, it possesses a null vector. Let us assume that  $L_{\dot q \dot q} \neq 0$ which corresponds to the case \eqref{case1}. 
Then,  as shown in \eqref{ssdiK} and \eqref{ssO},  $K$  admits the null vector  $(1,-L_{\dot q \dot Q}/L_{\dot q \dot q})$, which is a function of $(\dot Q,Q,\phi,\dot q,q)$. 
As a consequence, \eqref{system1} is equivalent to the
following two equations:
\bea\label{system2}
\lambda = V -\frac{L_{\dot q \dot Q}}{L_{\dot q \dot q}} v \, , \qquad  L_{\dot q \dot q} \ddot q + L_{\dot q \dot Q} \ddot Q = v \, .
\eea
Note that the first of these equations does not contain  second derivatives and determines the variable $\lambda$. Since $\lambda=\pi$, the first equation of \eqref{system2} can be seen  as the Lagrangian version of the secondary  constraint \eqref{secondTheta},  which we obtained in the Hamiltonian analysis.

The equations of motion for $\phi$ and $q$ are provided by the second equation in \eqref{system2} and  $L_\phi=\dot \lambda$, where $\lambda$ is replaced by its expression in \eqref{system2}. To reduce these two equations to a second order system, we need to use explicitly the  constraint \eqref{case1}, which can be written as  $L_{\dot Q} = F(p,Q,q,\phi)$ with $p=L_{\dot q}$.
The derivatives of this constraint yield the following useful relations:
\bea\label{relationsFp}
&&L_{\dot Q Q} = F_p L_{\dot q Q} + F_Q \;,\qquad
L_{\dot Q q} = F_p L_{\dot q q} + F_q \;,\qquad
L_{\dot Q \phi} = F_p L_{\dot q \phi} + F_\phi \;, \cr
&&L_{\dot Q \dot Q} = F_p L_{\dot q \dot Q} \;,\qquad \qquad \;\;
L_{\dot Q \dot q} = F_p L_{\dot q \dot q} \,.
\eea
From \eqref{EOMLeq1} one can then express $\lambda$  in terms of $F$ as follows
\bea
\lambda= L_Q - \frac{dF}{dt} = L_Q - F_p L_q - F_Q \dot Q - F_q \dot q -F_\phi \dot\phi  \, .
\eea
Using  the above  relations, one finds 
\bea
\frac{\partial \dot \lambda}{\partial \ddot Q} &= & L_{\dot Q Q} - F_p L_{\dot Q q} - F_Q  =  F_p(L_{\dot q Q} - L_{\dot Q q})\,, \\
\frac{\partial \dot \lambda}{\partial \ddot q} &= &  L_{\dot q Q} - F_p L_{\dot q q} - F_q  =  L_{\dot q Q} - L_{\dot Q q} \, .
\eea
Thus, the equation of motion for $\phi$, i.e.\  $L_\phi=\dot \lambda$, takes the form
\bea\label{eomphi1}
(L_{\dot q Q} - L_{\dot Q q})(\ddot q + F_p \ddot Q) = w \, ,
\eea
where $w$ depends only on $(\ddot \phi,\dot \phi,\phi; \dot q, q)$. When $L_{\dot q Q}= L_{\dot Q q}$, the equation of motion for $\phi$ is $w=0$: it is second order in $\phi$
and does not depend on $\ddot q$. When $L_{\dot q Q} \neq L_{\dot Q q}$, one can  combine \eqref{eomphi1} with the equation of motion for $q$ in \eqref{system2}, 
which can be written as
\be
 L_{\dot q \dot q} (\ddot q +F_p \ddot Q) = v\,,
\ee
to obtain an equation of motion for $\phi$ of the form
\be
\label{eomphi}
  {\cal E}(\ddot \phi,\dot \phi,\phi; \dot q, q)  \equiv L_{\dot q \dot q} w -  (L_{\dot q Q} - L_{\dot Q q}) v =  0\, ,
\ee
where ${\cal E}$ can be computed explicitly, although its expression is not simple in general.

As a consequence, the equation for $\phi$ is always a second order equation which involves  at most  the first derivative of $q$. Computing the time derivative of
this equation enables us to obtain generically (when $\partial {\cal E}/\partial \ddot \phi$ does not vanish) $\dddot \phi$ in terms of up to second derivatives of $q$ and $\phi$. Substituting this last
relation in the second equation of \eqref{system2} with $Q=\dot\phi$ leads to a second order equation for $q$ as well. This proves that the equations of motion can be recast as a second order system.
One can deal with the case \eqref{case2}  with an analogous procedure and reach  the same conclusions.

The remaining  case \eqref{case3} 
\beq L_{\ddot \phi} = f(\dot \phi,\phi,q), \qquad L_{\dot q} = g(\dot \phi,\phi,q)  \eeq
is simpler to analyze. Indeed, it is obvious that the fourth-order derivatives of $\phi$ and third-order derivatives of $q$ do not appear in the equation of motion \eqref{eom1} for $\phi$. Moreover, the terms with $\dddot\phi$ cancel as $L_{\dot \phi \ddot \phi}=f_{\dot \phi}$. Since the equation of motion \eqref{eom2} for $q$ in this case involves only up to second order derivatives, one thus concludes 
that the  Euler-Lagrange equations form  directly a second 
order system.

\medskip

In conclusion, degenerate Lagrangians of the form (\ref{Lagrangian}) are such that their equations of motion can be reformulated as a system of second order equations for $\phi$ and
$q$. This is consistent with the Hamiltonian analysis which shows that there is no extra degree of freedom in these theories. Nonetheless,  it is worth stressing that the Euler-Lagrange
equations derived from the Lagrangian in general do not give  directly  the ``minimal'' system of equations because they can involve up to fourth-order derivatives of $\phi$, as we saw. Demanding the Euler-Lagrange equations to be second order  is clearly not a necessary requirement in order to avoid the Ostrogradsky ghost.

\subsection{Examples of degenerate theories}%%%%%%%%%%%%%%%%%%%%
\label{ssec:ssexam}

To illustrate our previous considerations, we now give some concrete examples of degenerate theories of the form (\ref{Lagrangian}).

\emp{Example 1: Linear primary constraint}

One can construct degenerate Lagrangians by assuming that the function $F$ appearing in the primary constraint \eqref{case1} depends linearly on the momentum $p$, i.e.\
\beq
F(p,Q, \phi, q)=a(Q,\phi,q) p+b(Q,\phi,q)\,.
\eeq
In this case, it is easy to see that the corresponding Lagrangians  are of the form 
\bea
L_{eq}^{(1)} (\dot Q, Q; \dot \phi, \phi ; \dot q , q) = L_0(\dot q + a \dot Q;Q,q,\phi) + b\dot Q \,
\eea
where $L_0$ is arbitrary.  This is a special case of the toy model considered in Sec.~4 of \cite{Motohashi:2015pra}.

\emp{Example 2: Factorized Lagrangians}

Another class of examples is given by  Lagrangians, whose dependence on $\dot Q$ and $\dot q$ is factorized, i.e.\ of the form
\bea\label{factorized}
L_{eq}^{(1)} = L_1(\dot Q; Q,q,\phi) \, L_2(\dot q;Q,q,\phi) \,.
\eea
Such a Lagrangian leads to a primary constraint (\ref{case1}) with  $F(t)=at^\alpha$, where $\alpha \neq 1$
and $a$ a nonvanishing function $a(Q,\phi,q)$, if the functions $L_1$ and $L_2$ satisfy the differential equations
\bea
\frac{\partial L_1}{\partial \dot Q} = b \, L_1^{\alpha} \;\;\; \text{and} \;\;\; 
\left(\frac{\partial L_2}{\partial \dot q}\right)^\alpha = \frac{b}{a} L_2 \;,
\eea 
where $b=b(Q,\phi,q)$. 

Assuming $b/a > 0$ for simplicity, one can write explicitly $L_1$ and $L_2$ as 
\bea
L_1= [ (1-\alpha) b \, \dot Q + c_1]^{\frac{1}{1-\alpha}} \;\;\; \text{and} \;\;\;
L_2=\left[\frac{\alpha -1}{\alpha} \left( \frac{b}{a}\right)^{\frac{1}{\alpha}} \dot q + c_2\right]^{\frac{\alpha}{\alpha -1}}
\eea
where $c_1$ and $c_2$ are functions of $(Q,\phi,q)$ only. Choosing for instance $\alpha=2$, $c_1=-b=-4a=2$, and $c_2=0$, we obtain the  Lagrangian 
\bea
L = \frac{1}{2} \frac{\dot q^2}{1+\ddot \phi} \,,
\eea
whose Euler-Lagrange equations can be rearranged into $\dot{q} =
C(1+\ddot \phi)$ with $C$ being a constant. Similar Lagrangians have been considered in Sec.~7.1 of \cite{Gabadadze:2012tr}.

\emp{Example 3: Linear second derivative}

As an example for the case \eqref{case3}, we can consider 
\beq L=\ddot \phi\,  f(\dot \phi,\phi,q) + \dot q\,  g(\dot \phi,\phi,q). \eeq
We note that the terms involving $\dddot \phi$ in the Euler-Lagrange equations  vanish identically. 
However, when multiple variables of the type $\phi$ are considered, the Euler-Lagrange equations in general contain nonvanishing $\dddot \phi$ terms~\cite{Motohashi:2014opa}.
We will return to this point 
in Sec.~\ref{sec:special-only}.

\section{Lagrangian with multiple regular variables and single special variable}%%%%%%%%%%%%%%%%%%%%
\label{sec:single-multi}

We wish to extend our previous analysis  to multiple variables.  For pedagogical reasons, in Sec.~\ref{sec:single-multi} we start with theories that possess only one special variable $\phi$ along with multiple regular variables $q^i$.  In Sec.~\ref{sec:special-only}, we consider Lagrangians with multiple special variables only. 
In Sec.~\ref{sec:multi-multi}, we finally study the full generalization with both multiple special and regular variables.  In all cases, we assume that the regular subsystem, when present,  is by itself nondegenerate, i.e.\ their momenta are all independent. 
Although there are a few subtleties with multiple variables, the analysis in this section is very similar to the simpler case studied in Sec.~\ref{sec:single-single}. 
Under the  assumption that the $q^i$'s form a nondegenerate subsystem,
the degeneracy of the kinetic matrix will be shown to be a necessary and sufficient condition for getting rid of the Ostrogradsky ghost.   
The analysis of this section generalizes the degeneracy condition for the quadratic toy model  considered in \cite{Langlois:2015cwa} to general Lagrangians with a single special variable and multiple regular variables.

Let us consider a Lagrangian of the form
\beq
L(\ddot \phi,\dot \phi , \phi ; \dot q^i, q^i) \qquad (i=1,\cdots, m)\,.
\eeq 
As before, it is convenient, in particular for  the Hamiltonian analysis, to use the equivalent Lagrangian
\beq \label{Lag singlephi}
L_{eq}^{(1)} (\dot Q, Q; \dot \phi, \phi ; \dot q^i , q^i;\lambda)
\equiv L(\dot Q, Q, \phi ; \dot q^i , q^i) + \lambda (\dot \phi - Q)\,,
\eeq
which depends on $(m+3)$ variables.

Out of $(m+3)$ variables, the Lagrange multiplier $\lambda$ is clearly nondynamical and the corresponding DOF is automatically removed by the primary constraints as we shall see below. In general,  the theory  thus contains $(m+2)$ DOF.  In order to eliminate another  DOF, associated with the Ostrogradsky ghost, one needs additional constraints, which are provided  by degenerate Lagrangians. In this case, one ends up  with $(m+1)$ healthy DOF, which correspond to one DOF associated with the special variable $\phi$ and $m$ DOF associated with the regular variables $q^i$'s.
In that respect, the problem is very similar to the simpler case studied in Sec.~\ref{sec:single-single}.

\subsection{Constraints}%%%%%%%%%%%%%%%%%%%%
\label{ssec:consti}

As usual, we introduce the pairs of conjugate variables
\bea
\{ Q,P\} \; = \; \{\phi,\pi \} \; = \; \{\lambda,\rho \}  \; = \; 1 \,, \; \; \text{and} \;\;
 \{q^i,p_j \} \; =\delta_j^i \; .
\eea
The form of the Lagrangian \eqref{Lag singlephi} implies the existence of  two primary constraints
\beq \Phi = \pi -\lambda \approx 0 \;\;, \;\; 
\Psi = \rho \approx 0 \, . \eeq
If there is no further primary constraint, we can proceed exactly as in Sec.~\ref{ssec:ssHam1}.
In this way, we  find 2 second class constraints that reduce the $(m+3)$ initial DOF to $(m+2)$ DOF, one of which is
the Ostrogradsky ghost.

To obtain $(m+1)$ healthy DOF, we need additional  constraints, analogous to  \eqref{case1}--\eqref{case3}.   
These constraints must kill the ghost, but not a safe degree of freedom like one of the regular variables. To be certain that we will not eliminate one of 
the $q^i$'s variables, we assume, as  already emphasized in the introduction of this section,  that the subsystem of regular variables is nondegenerate, i.e.\ their momenta are all independent. More precisely, we assume that the relation defining the momenta
$p_i = {\partial L}/{\partial \dot q^i}$ is invertible and then one can locally express the velocities $\dot q^i$
in terms of the momenta $p_i$ (and of the remaining phase space variables). 
This requirement is equivalent to asking that the sub-kinetic matrix $L_{ij}$ defined by
\bea\label{nondegenerateq}
L_{ij} \equiv L_{\dot q^i \dot q^j}\equiv \frac{\partial^2 L}{\partial \dot q^i \partial \dot q^j}
\eea
is non degenerate. A consequence of this hypothesis is that only the  case \eqref{case1} can be generalized to  multiple regular variables. Thus, we look for
Lagrangians giving a primary constraint of the type
\beq \label{case1i} \Xi \equiv P-F(p_i, q^i,Q,\phi) \approx 0 , \eeq 
where $F$ is an arbitrary function. 
This condition is equivalent to  the degeneracy of the full $(m+1)$ dimensional kinetic matrix
\bea \label{fullKi}
K \equiv \begin{pmatrix}
L_{\dot Q \dot Q} & L_{\dot Q j} \\
L_{i \dot Q} & L_{ij}
\end{pmatrix}  \;\;\; \text{with} \;\;\; L_i \equiv \frac{\partial L}{\partial \dot q^i} \;\; \text{and } \;\; L_{i \dot Q } \equiv L_{\dot q^i \dot Q }\,.
\eea
Since the determinant of $K$ is given by 
\be
\det K = (L_{\dot Q \dot Q} - L_{\dot Q i} L^{ij} L_{j \dot Q}) \det L_{ij} 
\ee
where $L^{ij}L_{jk}=\delta^i_k$, the degeneracy of $K$, together with  $\det L_{ij} \neq 0$, implies
\be \label{degconi}
L_{\dot Q \dot Q} - L_{\dot Q i} L^{ij} L_{j \dot Q} = 0 .
\ee 
To prove the equivalence between \eqref{case1i} and \eqref{degconi}, one can use a strategy similar to that of  Sec.~\ref{sec:single-single}.

First, it is easy to show that \eqref{degconi} follows from \eqref{case1i}. Indeed,  \eqref{case1i} implies $L_{\dot Q} = F(L_i,Q,q^i,\phi)$ which in turn implies 
\bea \label{LQQLiQi}
L_{\dot Q \dot Q}= L_{i\dot Q} \frac{\partial F}{\partial p_i} \;\;\;\; \text{and} \;\;\;\;
L_{i\dot Q}= L_{ij} \frac{\partial F}{\partial p_j} \,.
\eea
To show the converse, one writes  the momentum in the form
\bea
P=F(p_i, q^i, \dot Q,Q,\phi) \,,
\eea 
where the  velocities $\dot q^i$ have been replaced by the momenta $p_i$, which is always possible to do since $L_{ij}$ is invertible.
If $\partial F/\partial \dot Q|_{p_i} \neq 0$, then one could express locally $\dot Q$ in terms of the momenta, which would mean that the Legendre transform $(\dot Q,\dot q^i) \mapsto (P,p_i)$
is invertible, in  contradiction with the degeneracy of $K$. Therefore, $F$ does not depend on $\dot Q$ and we obtain a primary constraint of the type \eqref{case1i}.
An alternative and more concrete proof of this equivalence is provided in Appendix~\ref{sec:explicit degeneracy}.

The Hamiltonian analysis of the theory closely follows that of Sec.~\ref{sec:single-single} and  we will not reproduce it  here.  It can be easily checked that  $\dot \Xi \approx 0$ generates a secondary constraint 
$\Theta \approx 0$. In general,  there is no further constraint and one ends up with 4 second class constraints. As a consequence, the theory admits generically $(m+1)$ DOF 
and there is no  Ostrogradsky ghost. In some particular cases, there may exist extra (tertiary) constraints and some of the constraints may be first class, as discussed in 
Sec.~\ref{sec:single-single}. In such cases, the theory could  possess only $m$ degrees of freedom, still without Ostrogradsky ghost.

\subsection{Euler-Lagrange equations}%%%%%%%%%%%%%%%%%%%%
\label{ssec:eomi}

To show that the Euler-Lagrange equations in degenerate theories reduce to a second order system, we follow the same strategy as in Sec.~\ref{ssec:ssELeq}. 
We first derive the equations of motion associated to the equivalent Lagrangian \eqref{Lag singlephi}:
\bea
\label{el1}
L_Q - \frac{d L_{\dot Q}}{dt} = \lambda \; , \qquad 
L_{q^i} - \frac{d L_{\dot q^i}}{dt}= 0 \; , \qquad  L_\phi = \dot \lambda \;,\qquad \dot \phi = Q \, .
\eea
To avoid confusion, we have returned in this section to the more explicit notation $L_{\dot q^i}$ instead of $L_i$ for $\partial L/\partial \dot q^i$.
The first two equations can be reformulated as
\bea\label{system1multiq}
K \left( \begin{array}{c} \ddot Q \\ \ddot q^i \end{array}\right) = 
\left( \begin{array}{c} V - \lambda \\ v_i \end{array}\right) \qquad \text{with} \;\; 
\left( \begin{array}{c} V \\ v_i \end{array}\right)=
\left( \begin{array}{c}  L_Q -L_{\dot Q Q} \dot Q -L_{\dot Q q^j} \dot q^j - L_{\dot Q \phi} \dot \phi \\  
L_{q^i}  -L_{\dot q^i Q} \dot Q - L_{\dot q^i q^j} \dot q^j - L_{\dot q^i \phi} \dot \phi \end{array}\right).
\eea
The kinetic matrix is degenerate in only one null direction defined by the vector $(-1,u^i)$ with
$u^i=L^{ij} L_{\dot q^j \dot Q}=\pa F/\pa p_i$, as shown in \eqref{diagki} and \eqref{smO}. Thus projecting the previous system in this null direction allows us to fix $\lambda$ to
\bea\label{lambdamultiq}
\lambda = V - u^i v_i \,.
\eea
The equations of motion for $Q$ and $q^i$ are given by
\bea \label{eomqimultiq}
L_{ij} \ddot q^j + L_{\dot q^i \dot Q} \ddot Q = v_i \;, \qquad \; L_\phi=\dot \lambda \, .
\eea
They involve a priori the third derivative of $\phi$. To get rid of these higher derivatives we make use of the primary constraint $L_{\dot Q}=F(p_i,q^i,Q,\phi)$ with $p_i=L_{\dot q^i}$. 
Furthermore, the expression of $\lambda$ \eqref{lambdamultiq} simplifies to
\bea
\label{el2}
\lambda = L_Q - F_{p_i} L_{q^i} - F_Q \dot Q -F_{q^i} \dot q^i - F_\phi \dot \phi  \,,
\eea
which gives
\bea
\frac{\partial \dot \lambda}{\partial \ddot Q} &= & L_{\dot Q Q} - F_{p_i} L_{\dot Q q^i} - F_Q  = F_{p_i} (L_{\dot q^i Q}-L_{ q^i \dot Q}) \;, \\
\frac{\partial \dot \lambda}{\partial \ddot q^i}&= & L_{\dot q^i Q} - F_{p_j} L_{\dot q^i q^j} - F_{q^i} = L_{\dot q^i Q}-L_{q^i \dot Q } +
F_{p_j} ( L_{\dot q^j q^i}-L_{q^j \dot q^i})\, ,
\eea
where we used relations similar to \eqref{relationsFp}.
As a consequence, the equation of motion for $\phi$, i.e.\ $L_\phi = \dot \lambda$,  takes the form
\bea\label{eomphimultiq}
(L_{\dot q^i Q} - L_{q^i \dot Q })(\ddot q^i + F_{p_i} \ddot Q) 
+F_{p_j} ( L_{\dot q^j q^i}-L_{q^j \dot q^i}) {\ddot q^i}
= w \, ,
\eea
where $w$ depends only on $(\ddot \phi,\dot \phi,\phi; \dot q^i, q^i)$.
One can  combine \eqref{eomphimultiq} with the equation of motion for $q^i$ in \eqref{eomqimultiq}, 
which can be written as
\be
 \ddot q^i = L^{i j} v_j - F_{p_i} \ddot Q \,,
\ee
to obtain an equation of motion for $\phi$ of the form
\be \label{eomphii}
{\cal E}(\ddot \phi,\dot \phi,\phi; \dot q^i, q^i) \equiv   
\kk{ (L_{\dot q^i Q} - L_{q^i \dot Q})
+F_{p_k} ( L_{\dot q^k q^i}-L_{q^k \dot q^i}) } L^{ij} v_j - w =  0\, ,
\ee
where ${\cal E}$ can be computed explicitly, although its expression is not simple in general.
Note that the coefficient for $\ddot Q=\dddot \phi$ vanishes identically.

We conclude that the equation for $\phi$ is always second order, and  involves at most  first derivatives of the $q^i$. Following the same reasoning as in the previous section, taking a time derivative of ${\cal E}$ allows us to write down $\dddot \phi$ as a function of terms up to second derivatives. Substituting this expression of $\dddot \phi$ into the first equation of \eqref{eomqimultiq}, we obtain  second order equations for all the $q^i$'s variables.  Thus 
the degeneracy condition, with $L_{ij}$ invertible, implies that the equations of motions can be written as a second order system.

\section{Lagrangian with only special variables}%%%%%%%%%%%%%%%%%%%%%%%%%%%%%%%%%%%%%%%%%
\label{sec:special-only}

Before going to the general analysis for Lagrangians with arbitrary numbers of regular and special variables in Sec.~\ref{sec:multi-multi}, 
we discuss in this section the particular case of Lagrangians that depend only on special variables.
As pointed out in~\cite{Motohashi:2014opa}, there is a qualitative difference when we consider multiple special variables.
For $L(\ddot \phi^a, \dot \phi^a, \phi^a)$ with $a=1,\cdots, n$, the Euler-Lagrange equations are in general fourth order, 
\be \f{\pa^2L}{\pa\ddot \phi^a\pa\ddot \phi^b} \ddddot \phi^b = \text{(lower derivatives)}. \ee
If the matrix $\pa^2L/\pa\ddot \phi^a\pa\ddot \phi^b$ is nondegenerate, 
one can  multiply the above system by  its inverse matrix to obtain $n$ fourth order EOMs of the form
\be \ddddot \phi^a = \text{(lower derivatives)}\,, \ee
which require $4n$ initial conditions.
In other words, we have $2n$ DOF and
half of them  are  Ostrogradsky ghosts, associated with a linear dependence of the Hamiltonian on their canonical momenta.

If some of the eigenvalues of the matrix $\pa^2L/\pa\ddot \phi^a\pa\ddot \phi^b$ vanish, one can take particular  linear combinations of EOMs to eliminate some fourth order derivatives.
Let us now consider the maximally degenerate case for  which 
\be \label{firstcond} \f{\pa^2L}{\pa\ddot \phi^a\pa\ddot \phi^b} = 0\,. \ee 
In that situation, the Lagrangian takes necessarily  the form
\be
L = \sum_{a} \ddot \phi^a f_a (\dot \phi^b,\phi^b) + g (\dot \phi^b,\phi^b) \,,
\ee
where $f_a$ and $g$ are $(N+1)$ arbitrary functions of the fields $\phi^b$ and their velocities $\dot\phi^b$.
The highest derivative terms in the EOMs are then third order: 
\be 
E_{ab}  \dddot \phi^b = 
\text{(lower derivatives)} \;\;\; \text{with} \;\;\;
E_{ab} \equiv  \f{\pa^2L}{\pa\ddot \phi^a\pa\dot \phi^b} - \f{\pa^2L}{\pa\ddot \phi^b\pa\dot \phi^a} = 
 \f{\pa f_a}{\pa\dot \phi^b} - \f{\pa f_b}{\pa\dot \phi^a} . \ee
If $\det E \neq 0$, the system is essentially third order and cannot be reduced to a lower order system.  
One thus needs to specify $3n$ initial conditions and the system still suffers from the Ostrogradsky  instability~\cite{Motohashi:2014opa}. This is a simple illustration of the fact that the degeneracy is not a sufficient condition for 
eliminating the Ostrogradsky ghost when several special variables are present.

To circumvent this problem, a sufficient condition is to require 
\be \label{secondcond} E_{ab}=0. \ee
With the conditions  \eqref{firstcond} and \eqref{secondcond}, the Euler-Lagrange equations for the Lagrangian $L(\ddot \phi^a, \dot \phi^a, \phi^a)$ are second order and only $2n$ initial conditions are needed, i.e.\ only $n$ DOF are present.  
This can be seen immediately from the fact that \eqref{secondcond} implies the existence of a function 
$F(\dot\phi^a,\phi^a)$ such that $f_a=\partial F/\partial \dot \phi^a$, hence
\be
L= \frac{d F}{dt} - \sum_{a} \dot \phi^a \frac{\partial F}{\partial \phi^a}(\dot \phi^b,\phi^b) + g (\dot \phi^b,\phi^b) \, . 
\ee
As a consequence, the second order derivatives are removed from the Lagrangian.

To summarize, the first condition \eqref{firstcond} is the usual degeneracy condition and eliminates the fourth order derivative terms in EOMs.  The second condition \eqref{secondcond}  eliminates the third order derivative terms in EOMs. 
As one can see from its anti-symmetric nature, the second condition \eqref{secondcond} only applies when several special variables are present, which is consistent with Example~3 in Sec.~\ref{ssec:ssexam}.
It is straightforward to extend  this discussion to Lagrangians of the form  $L(\phi^{a(N)}, \cdots,\dot\phi^a, \phi^a)$, with derivatives of arbitrary order $N$. The equations of motion are then of order $2N$, unless the matrix ${\pa^2L}/{\pa\phi^{a(N)}\pa\phi^{b(N)}}$ is degenerate. If this matrix vanishes, the EOMs
become of order $(2N-1)$ in general and the system  still suffers from the Ostrogradsky instability. One needs extra conditions similar to \eqref{secondcond}  to get rid of the Ostrogradsky ghosts~\cite{Motohashi:2014opa}.

\section{Lagrangian with multiple regular and special variables}%%%%%%%%%%%%%%%%%%%%
\label{sec:multi-multi}

In this section, we consider the general case of Lagrangians containing both multiple special variables $\phi^a$ and multiple regular variables $q^i$,
\beq \label{lagmm}
L(\ddot \phi^a,\dot \phi^a , \phi^a ; \dot q^i, q^i) \qquad (a=1,\cdots, n; \,\, i=1,\cdots, m).
\eeq 
As in Secs.~\ref{sec:single-single} and \ref{sec:single-multi}, we assume that the 
regular subsystem is by itself nondegenerate\footnote{It is of course possible to consider systems where  the variables $q^i$ are also degenerate. The present analysis is  straightforward to extend  although it would be more involved in practice.}. 
It is also convenient to use  the  equivalent Lagrangian 
\beq \label{equivformmultiphi}
L_{eq}^{(1)} (\dot Q^a, Q^a; \dot \phi^a, \phi^a ; \dot q^i , q^i,\lambda_a)
\equiv L(\dot Q^a, Q^a, \phi^a ; \dot q^i , q^i) + \lambda_a (\dot \phi^a - Q^a) ,
\eeq
where the Lagrange multipliers $\lambda_a$  can be treated as new variables.

In general, the Lagrangian \eqref{lagmm}, or equivalently \eqref{equivformmultiphi}, describes $(2n+m)$ DOF, each special variable being associated to 2 DOF. Our goal will be to identify a subclass of Lagrangians that are free of Ostrogradsky ghosts, which implies that they should contain at most $(n+m)$ DOF.

\subsection{Hamiltonian analysis}%%%%%%%%%%%%%%%%%%%%

Canonical variables are defined  by the following nontrivial Poisson brackets 
\bea
\{ Q^a,P_b\} = \{\phi^a,\pi_b \} =  \{\lambda^a,\rho_b \}  = \;  \delta^a_b  \;\; , \;\;  \{q^i,p_j \} \; =\delta_j^i .
\eea
The Lagrangian induces   two sets of $n$ primary constraints
\beq \Phi_a = \pi_a -\lambda_a \approx 0 \;\;, \;\; 
\Psi_a = \rho_a \approx 0 \, , \eeq
which can be used  to eliminate the extra-variables $\lambda_a$ together with their momenta $\rho^a$. 
If there are no other primary constraints, one can follow the procedure already discussed in Secs.~\ref{sec:single-single} and \ref{sec:single-multi}, and one ends up with $(2n+m)$ DOF, among which $n$ are Ostrogradsky ghosts

In order to eliminate the Ostrogradsky ghosts, we now assume, generalizing 
the constraint \eqref{case1i}  for  a single special variable discussed in Sec.~\ref{sec:single-multi},
that there exist $n$ primary constraints  of the form
\beq \label{case1ai} \Xi_a \equiv P_a-F_a(p_i, q^i, Q^b,\phi^b) \approx 0 .  \eeq 
The total Hamiltonian is then given by
\bea
H_T = H + \mu^a \Phi_a + \nu^a \Psi_a + \xi^a \Xi_a\,
\;\;\; \text{with} \;\;\;
H= P_a \dot Q^a + \rho_a \dot \lambda^a + \pi_a\dot \phi^a + p_i \dot q^i - L_{eq}^{(1)} ,
\eea
where $\mu^a$,  $\nu^a$ and $\xi^a$ are Lagrange multipliers. 
Requiring the time invariance of the primary constraints ${\Phi}_a$ and ${\Psi}_a$, using $\{ \Psi_a, \Phi_b \}=\d_{ab}$, 
fixes the Lagrange multipliers $\mu^a$ and $\nu^a$,  in particular $\mu^a = 0$. 
And time invariance of the remaining primary constraints $\Xi_a$ leads to the following $n$ conditions
\bea\label{stability Xia}
\dot \Xi_a = \{ \Xi_a , H \}   + \xi^b \{\Xi_a,\Xi_b\} \approx 0 \,.
\eea
The status of this set of conditions depends on the
$n$-dimensional matrix $M$ whose entries are $M_{ab} \equiv
\{\Xi_a,\Xi_b\}$. 
 If $M$ is invertible,  all the
Lagrange multipliers $\xi^a$ are determined and there are no secondary constraints. In
this case, the system does not have sufficient number of constraints to eliminate the ghost DOF.

In order to get secondary constraints, $M$ must be degenerate.  
The simplest scenario to get rid of all the Ostrogradsky
ghosts is to require that the whole matrix $M$ vanishes\footnote{It would also be possible to have a nonvanishing matrix $M$, thus yielding fewer than  $n$ secondary constraints. The elimination of all the ghosts would then require the existence of a sufficient number of further (tertiary, etc)  constraints.}
\  
\beq
\label{Xiab} \{ \Xi_a, \Xi_b \} = - \frac{\partial F_a}{\partial Q^b} +
\frac{\partial F_b}{\partial Q^a}{+ \frac{\partial F_a}{\partial
q_i}\frac{\partial F_b}{\partial p_i} - \frac{\partial F_a}{\partial p_i}\frac{\partial F_b}{\partial q_i}}= 0.  
\eeq 
In that case, \eqref{stability Xia} implies the existence of $n$
secondary constraints 
\be \Theta_a \equiv \{ \Xi_a , H \} \approx 0 \, .  \ee 
These constraints  fix all the
momenta $\pi_a$ in terms of the $p_i$ and of the canonical coordinates.  Since $\Xi_a$
does not contain $\pi_a$, the set of secondary constraints $\Theta_a$
is independent of the set $\Xi_a$.  Now, 
these constraints are sufficient to eliminate all the ghost-like DOF.

If in addition the matrix
$\Delta$ with entries $\Delta_{ab}=\{\Theta_a,\Xi_b\}$ is invertible,
then the primary and secondary constraints are all second class and we
end up with exactly $(n+m)$ DOF as required. If $\det\Delta=0$, then there
may be tertiary constraints or there might be first class
constraints in the theory. Thus, Lagrangians with degenerate $\Delta$
have fewer than $(n+m)$ DOF and none of them is an Ostrogradsky ghost.
In conclusion, the conditions \eqref{case1ai} and \eqref{Xiab} are
sufficient to define ghost-free higher derivative Lagrangians with multiple special variables.

\subsection{Conditions for the Lagrangian to evade the Ostrogradsky instability}%%%%%%%%%%%%%%%%%%%%
\label{ssec:constai}

In analogy with the results of the previous sections, 
one can show that the condition \eqref{case1ai} is equivalent to the degeneracy of the $(n+m)$-dimensional kinetic matrix 
\beq 
\label{kmatai} 
K=
\begin{pmatrix}
L_{ab} & L_{aj} \\
L_{ib} & L_{ij}
\end{pmatrix} \,,
\eeq
where we use  the notations 
\bea
L_{ij} \equiv \frac{\partial^2 L}{\partial \dot q^i \partial  \dot q^j}  \;\; , \;\; 
L_{ab} \equiv \frac{\partial^2 L}{\partial \dot Q^a \partial \dot Q^b} \;\; , \;\;
L_{ia} \equiv \frac{\partial^2 L}{\partial \dot Q^a \partial \dot q^i} .
\eea

More precisely, the degeneracy must be of order $n$, i.e.\ $\text{dim(Ker} \, K) = n$, which can be expressed by the conditions 
\beq \label{degcon1ai}  
L_{ab} - L_{ai} L^{ij} L_{jb} = 0 \,. 
\eeq  
Indeed, the condition  $\text{dim(Ker} \, K) = n$ 
is equivalent to the existence of $n$ eigenvectors $(v_\alpha^b,v_\alpha^i)$ for $\alpha \in \{1,\cdots, n \}$ such that
\beq 
\begin{pmatrix}
L_{ab} & L_{aj} \\
L_{ib} & L_{ij}
\end{pmatrix}  
\left(
\begin{array}{c}
v_\alpha^b \\
v_\alpha^j
\end{array}
\right) = 0 \quad \Longrightarrow  \quad (L_{ab} - L_{ai} L^{ij} L_{jb}) v_\alpha^b = 0 \,,
\eeq
where we have used the property that $L_{ij}$ is  invertible.
Since the $v_\alpha^b$ form a family of $n$ independent $n$-dimensional vectors, we conclude that $L_{ab} - L_{ai} L^{ij} L_{jb}=0$.  Conversely, if $L_{ab} - L_{ai} L^{ij} L_{jb}=0$, one can easily construct at least $n$ null-vectors of $K$, with their components   satisfying  $v_\alpha^i=-L^{ij} L_{jb} v_\alpha^b$. This, together with the invertibility of  $L_{ij}$,   implies $\text{dim(Ker} \, K) = n$. 

Let us  now   show the equivalence between  \eqref{case1ai} and  \eqref{degcon1ai}. 
It is immediate to see that  \eqref{case1ai} implies  \eqref{degcon1ai} by writing 
\bea
L_{ab} = L_{ib} \frac{\partial F_a}{\partial p^i}  \qquad \text{and} \qquad
L_{ia} = L_{ij} \frac{\partial F_a}{\partial p^j} , 
\eea
which directly follows from \eqref{case1ai}. The converse  is proved in a way similar to previous sections. As $L_{ij}$ is invertible, one can write any momentum
$P_a$ as a function $P_a=F_a(\dot Q^b,p_i,Q,q,\phi)$. If there exists  a pair $(a,b)$ such that {$\partial F_a/ \partial \dot Q^b|_{p_i} \neq 0$} then one sees immediately that 
$L_{ab} - L_{ai} L^{ij} L_{jb} = \partial F_a/ \partial \dot Q^b|_{p_i} \neq 0$. 
Thus the functions $F_a$ do not depend on the velocities $\dot Q^b$.   A more explicit proof is provided in Appendix~\ref{sec:explicit degeneracy}.

Finally, let us examine the consequences of  the  conditions  \eqref{Xiab}  for the Lagrangian. 
Taking derivatives of \eqref{case1ai} with respect to $Q^b$ and
$q^i$ with the use of \eqref{dfai}, we obtain 
\bea
&& \f{\pd F_a}{\pd Q^b} 
= \f{\pd^2 L}{\pd \dot Q^a \pd Q^b} - \f{\pd^2 L}{\pd \dot q^i \pd Q^b}
L^{ij}L_{aj}, \nonumber \\
&&
\f{\pd F_a}{\pd q^i} 
= \f{\pd^2 L}{\pd \dot Q^a \pd q^i} - \f{\pd^2 L}{\pd \dot q^j \pd q^i}
L^{jk}L_{ak}.  
\eea
Plugging these expressions together with \eqref{dfai} into \eqref{Xiab}
yields  the conditions
\bea \label{secdegcon1ai} 
0 & = &  \frac{\partial^2 L}{\partial \dot Q^a \partial \dot \phi^b}
- \frac{\partial^2 L}{\partial \dot Q^b \partial \dot \phi^a}
+ \frac{\partial^2 L}{\partial \dot \phi^a \partial \dot q^i} L^{ij} \frac{\partial^2 L} {\partial \dot q^j \partial \dot Q^b} 
- \frac{\partial^2 L}{\partial \dot Q^a \partial \dot q^i} L^{ij}
\frac{\partial^2 L} {\partial \dot q^j \partial \dot \phi^b}  \nonumber \\
&& 
+ \frac{\partial^2 L}{\partial \dot Q^a \partial \dot q^i} L^{ij} \frac{\partial^2 L} {\partial q^j \partial \dot Q^b} 
- \frac{\partial^2 L}{\partial \dot Q^a \partial q^i} L^{ij}
\frac{\partial^2 L} {\partial \dot q^j \partial \dot Q^b} \nonumber \\
&&
+ \frac{\partial^2 L}{\partial \dot Q^a \partial \dot q^i} L^{ij} 
\left(\frac{\partial^2 L} {\partial \dot q^j \partial q^k} 
    - \frac{\partial^2 L} {\partial q^j \partial \dot q^k} \right)
L^{kl} \frac{\partial^2 L} {\partial \dot q^l \partial \dot Q^b} 
 ,
\eea
where we have explicitly written some second derivatives of $L$ with respect to velocities to avoid confusion. 
The converse is also true.
Note that the above conditions reduce to \eqref{secondcond} in the absence of regular variables.

In conclusion, any Lagrangian of the form \eqref{lagmm} which satisfies the relations \eqref{degcon1ai} and \eqref{secdegcon1ai}
is  free of Ostrogradsky ghosts. These conditions have a clear interpretation from the Hamiltonian point of view: they ensure the existence of
primary and secondary second class constraints which enable one to get rid of the Ostrogradsky ghosts.

\subsection{Euler-Lagrange equations}%%%%%%%%%%%%%%%%%%%%
\label{ssec:eomai}

We conclude our study of multi-variable Lagrangians by  showing that the equations of motion can be written as a second order system. 
We follow the same method as in previous simpler cases starting from the equivalent formulation \eqref{equivformmultiphi} of Lagrangian.  
Euler-Lagrange equations can be written as
\bea
K \left( \begin{array}{c} \ddot Q^a \\  \ddot q^i\end{array} \right) = \left( \begin{array}{c} V_a - \lambda_a \\  v_i\end{array} \right) \; , \qquad
L_{\phi^a} = \dot \lambda_a \; , \qquad Q^a = \dot \phi^a ,
\eea
where $K$ is the kinetic matrix \eqref{kmatai} and $(V_a,v_i)$ is given by 
\bea
\left( \begin{array}{c} V_a \\ v_i \end{array}\right)=
\left( \begin{array}{c}  L_{Q^a} -L_{\dot Q^a Q^b} \dot Q^b -L_{\dot Q^a q^j} \dot q^j - L_{\dot Q^a \phi^b} \dot \phi^b \\  
L_{q^i}  -L_{\dot q^i Q^b} \dot Q^b - L_{\dot q^i q^j} \dot q^j - L_{\dot q^i \phi^b} \dot \phi^b \end{array}\right),
\eea   
which is written down in terms of up to the second order derivatives of $q$ and $\phi$.
As previously, $n$ among these equations fix the Lagrange multipliers $\lambda_a$ and they correspond to the secondary  constraints $\Phi_a$ in the Hamiltonian analysis.  The equations of motion for $\phi^a$ and $q^i$ take the form
\bea\label{eommultiall}
L_{\phi^a} = \dot \lambda_a  \;, \qquad L_{\dot Q^a \dot q^i} \ddot Q^a + L_{ij} \ddot q^j = v_i \; ,
\eea
where $\lambda_a$ is replaced by its expression $\lambda_a(\dot Q^b,Q^b,\phi^b;\dot q^i,q^i)$ (see Appendix~\ref{ssec:ed3}).

To go further, we first make use of the primary  constraints $L_{\dot Q^a}=F_a(p_i,q^i;Q^b,\phi^b)$ with $p_i=L_{\dot q^i}$.
Using these constraints, a straightforward calculation shows that the terms
proportional to $\ddot Q^b$ and $\ddot q^i$ in the equation of motion for $\phi^a$ are given by
\bea
\frac{\partial \dot \lambda_a}{\partial \ddot Q^b}  & = & L_{Q^a \dot Q^b} - \frac{\partial F_a}{\partial p_i} L_{q^i \dot Q^b} - \frac{\partial F_a}{\partial Q^b} \, ,\\
\frac{\partial \dot \lambda_a}{\partial \ddot q^i}  & = & L_{Q^a \dot q^i} - \frac{\partial F_a}{\partial p_j} L_{q^j \dot q^i} - \frac{\partial F_a}{\partial q^i} \, .
\eea
The equations of motion \eqref{eommultiall} then read
\bea
&&\frac{\partial \dot \lambda_a}{\partial \ddot Q^b}  \ddot Q^b + 
\frac{\partial \dot \lambda_a}{\partial \ddot q^i} \ddot q^i + {\cal R}_a = 0 \, , \label{eq1multi}\\
 && \ddot q^i = L^{ij} v_j - L^{ij} L_{\dot Q^b \dot q^j} \ddot Q^b \, , \label{eq2multi}
\eea
where ${\cal R}_a$ depends only on $(\ddot \phi^b,\dot \phi^b,\phi^b ; \dot q^j,\dot q^j)$. 
After substituting \eqref{eq2multi} into \eqref{eq1multi}, an immediate calculation shows that the coefficients of the $\ddot Q^b$ in the resulting equations are given by
\bea \label{sbsqdd}
\frac{\partial \dot \lambda_a}{\partial \ddot Q^b} - L^{ij} L_{\dot q^j \dot Q^b} \frac{\partial \dot \lambda_a}{\partial \ddot q^i} =
- \frac{\partial F_a}{\partial Q^b} + \frac{\partial F_b}{\partial Q^a}{+ \frac{\partial F_a}{\partial
q_i}\frac{\partial F_b}{\partial p_i} - \frac{\partial F_a}{\partial p_i}\frac{\partial F_b}{\partial q_i}} \, .
\eea
We recognize in the r.h.s.\ the Poisson brackets $\{ \Xi_a,\Xi_b\}$ between the secondary constraints $\Xi_a$. These coefficients are in general nonvanishing, in  contrast to the previous cases where the coefficient for $\ddot Q$ vanishes identically (see\eqref{eomphi} or \eqref{eomphii}).  
This illustrates the role of the extra conditions \eqref{Xiab} at the level of the equations of motion. Imposing them ensures that \eqref{sbsqdd} vanishes and that the $\ddot Q^b=\dddot \phi^b$ terms can be removed from the equations of motion for the $\phi^a$. We thus obtain
\bea
{\cal E}_a(\ddot \phi^b,\dot \phi^b,\phi^b ; \dot q^j,\dot q^j) \equiv  
\frac{\partial \dot \lambda_a}{\partial \ddot q^i} L^{ij} v_j + {\cal R}_a 
=0 \, .
\eea
Following the same idea as in the previous cases, and taking a time derivative of ${\cal E}_a$, we can write down $\dddot \phi^a$ in terms of up to second derivatives.  Plugging it into \eqref{eq2multi}, we obtain a set of second order equations of motion for $q^i$'s.  This concludes our analysis.

\section{Conclusion}%%%%%%%%%%%%%%%%%%%%%%%%%%%%%%%%%%%%%%%%
\label{sec:conc}

In this work, we have investigated in which circumstances  a classical mechanics Lagrangian containing  higher order derivatives can escape the generic Ostrogradsky instability.  We have shown that there is a qualitative difference between Lagrangians that contain only one special variable and those with multiple special variables.

In the first case,  the degeneracy of the kinetic matrix is a necessary and sufficient condition (under the assumption that the regular variables $q^i$ form a nondegenerate subsystem) to evade the Ostrogradsky instability.  The degeneracy of the kinetic matrix is associated with 
the existence of a primary constraint in  phase space, whose time invariance implies a secondary constraint. Both constraints eliminate the would-be Ostrogradsky ghost. 
This result holds for any number of regular variables and the degeneracy is expressed by simple conditions on the second derivatives of the Lagrangian (see Eq.~\eqref{degconi}).

By contrast, when $n (>1)$ special variables are present, the degeneracy, of order $n$, of the kinetic matrix (expressed by the conditions \eqref{degcon1ai})  is not sufficient to eliminate the $n$ Ostrogradsky ghosts that are present in general. The reason is that the degeneracy of order $n$ induces $n$ primary constraints, but requiring the time invariance of these constraints does not necessarily generate $n$ secondary constraints. Therefore, the degeneracy condition is not sufficient in general to get rid of the Ostrogradsky instability. This can however be achieved by imposing additional conditions, such as the vanishing of all  Poisson brackets between the primary constraints, which leads to the presence of  $n$ secondary constraints. In the Lagrangian formulation, these conditions can be expressed  as antisymmetric relations between the second derivatives of the Lagrangian with respect to the second or first order time derivatives of the various variables (see Eq.~\eqref{secdegcon1ai}).

In all cases, we showed  how  the higher order Euler-Lagrange equations can be rewritten as a second-order system. We also provided some specific examples of ghost-free Lagrangian (see Sec.~\ref{ssec:ssexam}). Although our results apply to Lagrangians describing point particles, we believe  that the conditions obtained in this paper  could be quite useful to construct ghost-free field theories involving for example several scalar fields and other fields such as the gravitational metric. 
It would thus be interesting to extend the present analysis  to field theories, a task which we leave for a future work.

\acknowledgements{%%%%%%%%%%%%%%%%%%%%%%%%%%%%%%%%%%%%%%%%%
We thank the organizers and participants of the workshop ``Exploring Theories of Modified Gravity'' at Kavli Institute for Cosmological Physics at University of Chicago (October 2015), where this work was initiated.  
This work was supported in part by 
Japan Society for the Promotion of Science (JSPS) Grant-in-Aid for Young Scientists (B) No.\ 15K17632 (T.S.), 
JSPS Grant-in-Aid for Scientific Research Nos.\ 25287054 (M.Y.), 26610062 (M.Y.),
MEXT Grant-in-Aid for Scientific Research on Innovative Areas ``New Developments in Astrophysics Through Multi-Messenger Observations of Gravitational Wave Sources'' Nos.\ 15H00777 (T.S.) and ``Cosmic Acceleration'' No.\ 15H05888 (T.S. \& M.Y.).
}

\appendix%%%%%%%%%%%%%%%%%%%%%%%%%%%%%%%%%%%%%%%%

\section{Expression for $\Delta$}%%%%%%%%%%%%%%%%%%%%%%%%%%%%%%%%%%%%%%%%
\label{sec:delta}

In this Appendix we derive the expression of $\Delta \equiv \{ \Theta, \Xi \}$, where $\Xi$ and $\Theta$ are defined in \eqref{case1} and \eqref{secondTheta}, respectively.  
Let us start by expressing the secondary constraint $\Theta$ as
\be
\Theta = -\pi+L_Q-F_Q {\dot Q}
-F_\phi Q -F_q {\dot q}-
F_p L_q. \label{ex-theta}
\ee
Even if velocities $\dot Q$ and $\dot q$ seem to enter in this expression, it is simple to show that $\Theta$ is a function of the phase space variables only.

Now, we provide how to obtain $\Delta$ given in \eqref{ex-Delta2} expressed by derivatives of Lagrangian.
First, it is straightforward to write down
\begin{align}
\Delta &= \{ \Theta, P-F \} \nonumber \\
&=\Theta_Q+\Theta_P F_Q+
\Theta_\pi F_\phi -
\Theta_q F_p+
\Theta_p F_q. \label{ex-Delta}
\end{align}
In order to proceed further, we need to know how $\Theta$ changes 
under the infinitesimal variation of the canonical variables.
To this end, let us perturb \eqref{ex-theta}.
The result is given by
\begin{align}
\delta \Theta =& L_{Q {\dot Q}} \delta {\dot Q}+L_{QQ} \delta Q +L_{Q {\dot q}} \delta {\dot q}
+L_{Qq} \delta q +L_{Q \phi} \delta \phi -\delta \pi \nonumber \\
&-{\dot Q} \left( F_{Qp} \delta p+F_{QQ} \delta Q
+F_{Q\phi} \delta \phi+F_{Qq} \delta q \right)
-F_Q \delta {\dot Q} \nonumber \\
&-Q \left( F_{\phi p} \delta p+F_{\phi Q} \delta Q
+F_{\phi \phi} \delta \phi+F_{\phi q} \delta q \right)
-F_\phi  \delta Q \nonumber \\
&-{\dot q} \left( F_{qp} \delta p+F_{qQ} \delta Q
+F_{q\phi} \delta \phi+F_{qq} \delta q \right)
-F_q \delta {\dot q} \nonumber \\
&-L_q \left( F_{pp} \delta p+F_{pQ} \delta Q
+F_{p\phi} \delta \phi+F_{pq} \delta q \right) \nonumber \\
&-F_p \left( L_{q\dot Q} \delta {\dot Q}+L_{qQ} \delta Q
+L_{q\dot q} \delta {\dot q}+L_{qq} \delta q 
+L_{q\phi} \delta \phi \right). \label{delta-theta}
\end{align}
Picking up velocity variation part only, we have
\be
\delta \Theta = \left( L_{Q {\dot Q}} -F_Q -F_p L_{q {\dot Q}} \right) \delta {\dot Q}+
\left( L_{Q {\dot q}} -F_q -F_p L_{q {\dot q}} \right) \delta {\dot q}
+\cdots. \label{velocity-variation}
\ee
Using the primary constraint \eqref{case1} written in the Language of the Lagrangian formalism, 
$L_{\dot Q}=F(L_{\dot q},Q,q,\phi)$, 
and definition of the conjugate momenta, \eqref{velocity-variation} becomes
\begin{align}
\delta \Theta 
&= (L_{Q {\dot q}}-L_{{\dot Q}q}) \mk{ F_p \delta \dot Q + \delta \dot q } +\cdots \notag\\
&= \frac{L_{Q {\dot q}}-L_{{\dot Q}q} }{L_{{\dot q}{\dot q}}}
(\delta p-L_{{\dot q}Q} \delta Q -L_{{\dot q}q} \delta q-L_{{\dot q}\phi} \delta \phi) +\cdots, \label{delTh}
\end{align}
where we used \eqref{dpeigenmode} and $L_{\dot q \dot q}\neq 0$.
As it should be from the fact that $\Theta$ is a function of the canonical variables,
$\delta \Theta$ has been finally expressed as a linear combination of the infinitesimal variation of the canonical variables.
Then, we find
\begin{eqnarray}
\Theta_Q&=&L_{QQ}-{\dot Q} F_{QQ}-Q F_{\phi Q}-F_\phi -{\dot q} F_{qQ}-
L_q F_{pQ}-F_p L_{qQ}
-(L_{Q{\dot q}}-L_{{\dot Q}q}) \frac{L_{{\dot q}Q}}{L_{{\dot q}{\dot q}}}, \nonumber \\
\Theta_P&=&0, \nonumber \\
\Theta_\pi &=&-1, \label{derivTh} \\
\Theta_q&=& L_{Qq}-{\dot Q} F_{Qq}
-Q F_{\phi q}-{\dot q} F_{qq}-
L_q F_{pq}-F_p L_{qq}
-(L_{Q{\dot q}}-L_{{\dot Q}q}) \frac{L_{{\dot q}q}}{L_{{\dot q}{\dot q}}}, \nonumber \\
\Theta_p&=&-{\dot Q} F_{Qp}-Q F_{\phi p}
-{\dot q} F_{qp}-L_q F_{pp}
+\frac{L_{Q{\dot q}}-L_{{\dot Q}q}}{L_{{\dot q}{\dot q}}}. \nonumber
\end{eqnarray}

It is appropriate to make one remark here.
Although $\Theta$ is a function of the canonical variables,
its specification is not unique in the sense that there is ambiguity of expressing
$\Theta$ in terms of the canonical variables due to the constraint $\Xi \approx 0$.
For instance, it is always possible to replace all $P's$ appearing in $\Theta$ by other 
variables by using $P=F(p,Q,q,\phi)$.
By the same token, it is equally allowed to partially keep $P$ in $\Theta$. 
This ambiguity amounts to adding $\alpha \delta \Xi$ to $\delta \Theta$ where $\alpha$ is an arbitrary function of the canonical variables.
Indeed, we can derive the following relation;
\begin{align}
\delta \Theta+\alpha \delta \Xi &=\delta \{ \Xi, H \}+\alpha \delta \Xi \nonumber \\
&=\delta ( \{ \Xi, H\}+\alpha \Xi ) -  \Xi \delta \alpha \nonumber \\
&=\delta \{ \Xi, H+\alpha' \Xi \} - \Xi \delta \alpha \nonumber \\
&\approx \delta \{ \Xi, H+\alpha' \Xi \},
\end{align}
where $\alpha = \{ \Xi, \alpha' \}$.
This shows that adding $\alpha \delta \Xi$ is equivalent to replace $P$
in $H$ by $F(p,Q,q,\phi)$ by some amount controlled by $\alpha'$.

What remains is to express the derivatives of $F$ in \eqref{ex-Delta} and \eqref{derivTh}
in terms of derivatives of Lagrangian.
For the sake of clarity, let us write the primary constraint as
\be
L_{\dot Q}=F(L_{\dot q},x^i),~~~~~~x^1=Q,~x^2=q,~x^3=\phi.
\ee
Taking the first derivative we obtain 
\be
\frac{\pa F}{\pa x^i}=\frac{\pa L_{\dot Q}}{\pa x^i}-F_p \frac{\pa L_{\dot q}}{\pa x^i},
\ee
and the second derivative yields
\be
\frac{\pa^2 F}{\pa x^i \pa x^j}=
\frac{\pa^2 L_{\dot Q}}{\pa x^i \pa x^j}-
\frac{\pa L_{\dot q}}{\pa x^i}\frac{\pa L_{\dot q}}{\pa x^j} F_{pp}
-\left( \frac{\pa L_{\dot q}}{\pa x^i} \frac{\pa F_p}{\pa x^j}+
\frac{\pa L_{\dot q}}{\pa x^j} \frac{\pa F_p}{\pa x^i} \right)
-\frac{\pa^2 L_{\dot q}}{\pa x^i \pa x^j} F_p.
\ee
In a similar way, we obtain
\begin{align}
\frac{\pa F_p}{\pa x^i}&=\frac{1}{L_{\dq \dq}}
\left( \frac{\pa L_{\dQ \dq}}{\pa x^i}-F_{pp} L_{\dq \dq}
\frac{\pa L_{\dq}}{\pa x^i}-F_p 
\frac{\pa L_{\dq \dq}}{\pa x^i} \right), \notag \\
F_{pp}&=\frac{1}{L_{\dq \dq}^2}
\left( L_{\dQ \dq \dq}-F_pL_{\dq \dq \dq} \right), \\
F_p&=\frac{L_{\dQ \dq}}{L_{\dq \dq}}. \notag
\end{align}
Plugging the derived expressions into \eqref{ex-Delta}, we finally obtain the following expression for $\Delta$:
\begin{eqnarray}
L_{\dq \dq}^3 \Delta &=&-{\dot m} \varepsilon^{ab\phi} \varepsilon^{cd\phi}
\delta^{\alpha \beta}_{m \dq} \delta^{\gamma \delta}_{d \dq} L_{\dq {\dot a}}
L_{\dq \alpha} L_{{\dot c}\gamma} L_{\delta \beta {\dot b}}
+\varepsilon^{ab\phi} \varepsilon^{cd\phi} \delta^{\alpha \beta}_{d\dq}
L_q L_{\dq {\dot a}} L_{{\dot c}\alpha} L_{{\dot b}\beta \dq} \nonumber \\
&&-2 \varepsilon^{ab\phi} L_{\dq \dq}^2 L_{\dq {\dot a}} L_{{\dot b}\phi}
+2 \delta^{\alpha \beta}_{q\dq} L_{\dq \dq}^2 L_{\dQ \alpha} L_{\beta Q}
+L_{\dq \dq}^2 (L_{\dQ \dQ}L_{qq}+L_{\dq \dq}L_{QQ}). \label{ex-Delta2}
\end{eqnarray}
Here, Einstein summation convention is used and the Roman/Greek letters denote $\{q,Q,\phi \}$/ $\{ q,Q,\phi,\dq,\dQ,{\dot \phi}\}$ respectively.
The $\varepsilon^{abc}$ is the totally anti-symmetric matrix with
$\varepsilon^{qQ\phi}=1$ and the generalized Kronecker delta 
$\delta^{\alpha \beta}_{\gamma \delta}$ is defined by
$\delta^{\alpha \beta}_{\gamma \delta} = \delta^\alpha_\gamma \delta^\beta_\delta
-\delta^\alpha_\delta \delta^\beta_\gamma$.

\section{Degeneracy of kinetic matrix and primary constraints}%%%%%%%%%%%%%%%%%%%%%%%%%%%%%%%%%%%%%%%%
\label{sec:explicit degeneracy}

This Appendix is devoted to showing more explicitly that the degeneracy of the kinetic matrix leads to the existence of primary constraints (see Secs.~\ref{ssec:ssdeg}, \ref{ssec:consti}, and \ref{ssec:constai}). 
We study 
the cases with single regular and special variables, 
multiple regular variables and single special variable,
and multiple regular and special variables.

\subsection{Single regular and special variables}%%%%%%%%%%%%%%%%%%%%
\label{ssec:ed1}

We provide an alternative and more concrete proof of $\det K=0$, where $K$ is defined in \eqref{kinmat}, leads to one of the three additional primary conditions \eqref{case1}, \eqref{case2} or \eqref{case3}.
For that purpose, we start writing the degenerate kinetic matrix as follows
\beq \label{formK} K = 
\begin{pmatrix}
a & b \\
b & c
\end{pmatrix} \quad \text{with} \quad ac =b^2 .\eeq  
Its degeneracy implies that one or two of its eigenvalues are vanishing.

Furthermore, $K$ is a real symmetric matrix and thus is diagonalizable. The explicit diagonalization depends on whether $c$ and $a$ are vanishing or not. 
First, if $c\equiv L_{\dot q \dot q}\neq 0$, $K$ has one nonzero eigenvalue and can be diagonalized as
\beq \label{ssdiK} K = 
\begin{pmatrix}
cr^2 & b \\
b & c
\end{pmatrix} 
= 
O^T
\begin{pmatrix}
0 & 0 \\
0 & c(r^2+1)
\end{pmatrix} 
O , \eeq  
where $r\equiv b/c$ and 
\beq \label{ssO} O = 
\frac{1}{\sqrt{r^2+1}}
\begin{pmatrix}
-1 & r \\
r & 1
\end{pmatrix}
=O^T=O^{-1} .
\eeq 
Now, we are going to show that the degeneracy leads in this case to a constraint \eqref{case1}. 
Indeed, using this eigenbasis of $K$ in \eqref{dpddq} leads to
\beq  
\begin{pmatrix}
-\d P + r\d p \\
r \d P + \d p
\end{pmatrix}
=
\begin{pmatrix}
0 \\
c(r^2 + 1)(r \d \dot Q + \d \dot q)
\end{pmatrix}  .
\eeq
We thus arrive at
\beq \label{dpeigenmode}
\d P - \frac{L_{\dot q \dot Q}}{L_{\dot q \dot q}} \d p = 0 , \qquad
\frac{L_{\dot q \dot Q}}{L_{\dot q \dot q}} \d \dot Q + \d \dot q = \frac{1}{L_{\dot q \dot q}} \d p.
\eeq
The function $L_{\dot q \dot Q}/L_{\dot q \dot q}$ is a priori a function of the velocities $\dot q$ and $\dot Q$. From the Legendre transform, it can be viewed as 
a function of $p$ and $\dot Q$ which in fact can be shown to depend only $p$ (and $Q,\phi,q$). Indeed, when one computes variations of $L_{\dot q \dot Q}/L_{\dot q \dot q}$
with respect to $\delta \dot q$ and $\delta \dot Q$ first and with respect to $\delta p$ and $\delta \dot Q$ using \eqref{dpeigenmode}, one obtains
\bea
\delta \left(\frac{L_{\dot q \dot Q}}{L_{\dot q \dot q}} \right) & = & \delta \dot q \frac{\partial }{\partial \dot q} \left( \frac{L_{\dot q \dot Q}}{L_{\dot q \dot q}}\right)
+ \delta \dot Q \frac{\partial }{\partial \dot Q} \left( \frac{L_{\dot q \dot Q}}{L_{\dot q \dot q}}\right) \nonumber \\
&=& \delta p \frac{1}{L_{\dot q \dot q}}  \frac{\partial }{\partial \dot q} \left( \frac{L_{\dot q \dot Q}}{L_{\dot q \dot q}}\right)  +
\delta \dot Q \frac{\partial }{\partial \dot Q} \kk{ \frac{L_{\dot q \dot Q}}{L_{\dot q \dot q}} - 
\frac{L_{\dot q \dot Q}}{L_{\dot q \dot q}}  \frac{\partial }{\partial \dot q} \left( \frac{L_{\dot q \dot Q}}{L_{\dot q \dot q}}\right) } \nonumber \\
&=& \delta p \frac{1}{L_{\dot q \dot q}}  \frac{\partial }{\partial \dot q} \left( \frac{L_{\dot q \dot Q}}{L_{\dot q \dot q}}\right)  +
\delta \dot Q \frac{1}{2L_{\dot q \dot q}^2} \frac{\partial }{\partial \dot q} \left( \text{det} K\right) \,. 
\eea
As $\text{det} K=0$, $L_{\dot q \dot Q}/L_{\dot q \dot q}$ is a function of $p$ only.
Thus,  the first equation in \eqref{dpeigenmode} gives the primary constraint (\ref{case1}) with $F'(p)= L_{\dot q \dot Q}/L_{\dot q \dot q}$.  
The second equation of \eqref{dpeigenmode}  is discussed in \eqref{delTh}.

Then, the case $a\equiv L_{\dot Q\dot Q}\neq 0$ in \eqref{formK} is treated in a way similar to the previous case, and leads to 
a primary constraint of the type \eqref{case2}. Finally, when $K$ has two vanishing eigenvalues, necessarily $K=0$, which leads immediately to constraints of the type \eqref{case3}.

\subsection{Multiple regular variables and single special variable}%%%%%%%%%%%%%%%%%%%%
\label{ssec:ed2}

We show that imposing the degeneracy condition \eqref{degconi} and $\det L_{ij} \neq 0$ to the kinetic matrix \eqref{fullKi} leads to the existence of the primary constraint \eqref{case1i}.
Using \eqref{degconi} and defining $u^i \equiv L^{ij} L_{j \dot Q}$, we can write 
\beq \label{Keli}
L_{i \dot Q } = L_{ij} u^j , \qquad
L_{\dot Q\dot Q} = L_{ij} u^i u^j  ,
\eeq
which amounts to \eqref{LQQLiQi}.
In fact, the $(n+1)$-dimensional vector $(-1,u^i)$ with $u^i=\pa F/\pa p_i$ is a null vector of $K$. We make use of this null vector to block-diagonalize $K$ as follows 
\beq \label{diagki}
K=
\begin{pmatrix}
L_{\dot Q \dot Q} & L_{\dot Q j} \\
L_{i\dot Q} & L_{ij}
\end{pmatrix} 
=
\begin{pmatrix}
u^T L u & u^T L \\
L u & L
\end{pmatrix} 
= 
T^{-1}
\begin{pmatrix}
0 & 0 \\
0 & C L C
\end{pmatrix} 
T ,
\eeq
where
\beq \label{smO} T = 
\begin{pmatrix}
(u^Tu+1)^{-1/2} & 0 \\
0 & C^{-1}
\end{pmatrix}
\begin{pmatrix}
-1 & u^T \\
u & 1
\end{pmatrix} , \quad
T^{-1} = 
\begin{pmatrix}
-1 & u^T \\
u & 1
\end{pmatrix}
\begin{pmatrix}
(u^Tu+1)^{-1/2} & 0 \\
0 & C^{-1}
\end{pmatrix} 
,
\eeq 
and an $m\times m$ matrix $C=(uu^T+1)^{1/2}$ is the square root of $(u^iu^j+\delta^{ij})$.
As the kinetic matrix relates the infinitesimal variations as $(\d P , \d p_i)^T = K (\d \dot Q , \d \dot q^j)^T$, evaluating it in the block-diagonalized basis yields
\beq  
\begin{pmatrix}
-\d P + u^T \d p \\
u \d P + \d p
\end{pmatrix}
=
\begin{pmatrix}
0 \\
(uu^T + 1) L (u \d \dot Q + \d \dot q)
\end{pmatrix}  .
\eeq
We thus arrive at
\beq \label{dpeigenmodei}
\d P - L_{\dot Q i} L^{ij} \d p_j = 0 , \qquad
L^{ij} L_{\dot Q j} \d \dot Q + \d \dot q^i = L^{ij} \d p_j ,
\eeq
which is precisely a generalization of \eqref{dpeigenmode} to the case with multiple regular variables.
We can confirm that the infinitesimal variation of $L^{ij} L_{\dot Q j}$ with respect to $\d \dot Q$ and $\d \dot q^i$ is given by
\beq \d(L^{ij} L_{\dot Q j}) 
= \d\dot Q \frac{\partial(L^{ij} L_{\dot Q j})}{\partial \dot Q} + \d\dot q^k \frac{\partial(L^{ij} L_{\dot Q j})}{\partial \dot q^k} 
= \d p_\ell L^{\ell k} \frac{\partial(L^{ij} L_{\dot Q j})}{\partial \dot q^k} . \eeq
Thus the first equation of \eqref{dpeigenmodei} gives the primary constraint \eqref{case1i} with $\partial F/\partial L_i= L^{ij} L_{\dot Q j}$.

\subsection{Multiple regular and special variables}%%%%%%%%%%%%%%%%%%%%
\label{ssec:ed3}

Similarly to Appendix~\ref{ssec:ed2}, imposing the degeneracy condition \eqref{degcon1ai} and $\det L_{ij} \neq 0$, the kinetic matrix \eqref{kmatai} can be block-diagonalized as
\beq 
K=
\begin{pmatrix}
L_{ab} & L_{aj} \\
L_{ib} & L_{ij}
\end{pmatrix} 
=
\begin{pmatrix}
A^T k A & A^T k \\
k A & k
\end{pmatrix} 
=
T^{-1}
\begin{pmatrix}
0 & 0 \\
0 & C k C
\end{pmatrix} 
T , \label{diagkai}
\eeq
where $A^i_a \equiv L^{ij} L_{ja}$, $k_{ij}\equiv L_{ij}$ to avoid confusion,
\beq \label{mmT}
T =
\begin{pmatrix}
B^{-1} & 0 \\
0 & C^{-1}
\end{pmatrix} 
\begin{pmatrix}
-1 & A^T \\
A & 1
\end{pmatrix} 
, \quad
T^{-1} =
\begin{pmatrix}
-1 & A^T \\
A & 1
\end{pmatrix} 
\begin{pmatrix}
B^{-1} & 0 \\
0 & C^{-1}
\end{pmatrix} ,
\eeq 
and an $n\times n$ matrix $B$ and an $m\times m$ matrix $C$ are the square roots of $A^TA+1$ and $AA^T+1$, respectively:
\beq B^2 = A^TA+1, \quad C^2 = AA^T+1  .  \eeq
Since all the eigenvalues for $A^TA+1$ and $AA^T+1$ are positive, $B$ and $C$ are well-defined.
Further, they are symmetric and have symmetric inverse matrices as their determinants are nonvanishing.
Substituting the block-diagonalization \eqref{diagkai} into the relation $(\d P_a , \d p_i)^T = K (\d \dot Q^b , \d \dot q^j)^T$, we obtain
\beq  
\begin{pmatrix}
-\d P + A^T \d p \\
A \d P + \d p
\end{pmatrix}
=
\begin{pmatrix}
0 \\
(AA^T+1) k (A \d \dot Q + \d \dot q)
\end{pmatrix}  .
\eeq
We thus arrive at
\beq \label{dpeigenmodeai}
\d P_a - L_{ai} L^{ij} \d p_j = 0 , \qquad
L^{ij} L_{aj} \d \dot Q^a + \d \dot q^i = L^{ij} \d p_j .
\eeq
which is a generalization of \eqref{dpeigenmode} or \eqref{dpeigenmodei}.
We can confirm that the infinitesimal variation of $L^{ij} L_{aj}$ with respect to $\d \dot Q$ and $\d \dot q^i$ is given by
\beq \d(L^{ij} L_{aj}) 
= \d\dot Q_b \frac{\partial(L^{ij} L_{aj})}{\partial \dot Q_b} + \d\dot q^k \frac{\partial(L^{ij} L_{aj})}{\partial \dot q^k} 
= \d p_\ell L^{\ell k} \frac{\partial(L^{ij} L_{aj})}{\partial \dot q^k} . \eeq
Thus the first equation of \eqref{dpeigenmodeai} gives the primary constraint \eqref{case1ai} with 
\beq \label{dfai} \f{\partial F_a}{\partial p_i}= L^{ij} L_{aj} .  \eeq

\bibliography{refs-degen}

\end{document}